\documentclass{aa}

\usepackage[english]{babel}
\usepackage[T1]{fontenc}
\usepackage[varg]{txfonts}
\usepackage{mathptmx}
\usepackage{amsmath}
\usepackage{amssymb}
\usepackage{amstext}
\usepackage{amsfonts}
\usepackage{mathrsfs}
\usepackage{graphicx}
\usepackage{float}
\bibpunct{(}{)}{;}{a}{}{,}
\newcommand{\mockalph}[1]{}

\begin{document}

\title{Initial mass function of planetesimals\\formed by the streaming instability}
\author{Urs Sch\"{a}fer{\inst{\ref{Hamburg},\ref{Lund}}}
\and Chao-Chin Yang\inst{\ref{Lund}}
\and Anders Johansen\inst{\ref{Lund}}}
\institute{Hamburg Observatory, University of Hamburg, Gojenbergsweg 112, 21029 Hamburg, Germany,\\\email{urs.schaefer@hs.uni-hamburg.de}\label{Hamburg}
\and Lund Observatory, Department of Astronomy and Theoretical Physics, Lund University, Box 43, 22100 Lund, Sweden\label{Lund}}
\date{}
\abstract{
The streaming instability is a mechanism to concentrate solid particles into overdense filaments that undergo gravitational collapse and form planetesimals. However, it remains unclear how the initial mass function of these planetesimals depends on the box dimensions of numerical simulations. To resolve this, we perform simulations of planetesimal formation with the largest box dimensions to date, allowing planetesimals to form simultaneously in multiple filaments that can only emerge within such large simulation boxes. In our simulations, planetesimals with sizes between 80 km and several hundred kilometers form. We find that a power law with a rather shallow exponential cutoff at the high-mass end represents the cumulative birth mass function better than an integrated power law. The steepness of the exponential cutoff is largely independent of box dimensions and resolution, while the exponent of the power law is not constrained at the resolutions we employ. Moreover, we find that the characteristic mass scale of the exponential cutoff correlates with the mass budget in each filament. Together with previous studies of high-resolution simulations with small box domains, our results therefore imply that the cumulative birth mass function of planetesimals is consistent with an exponentially tapered power law with a power-law exponent of approximately~$-1.6$ and a steepness of the exponential cutoff in the range of~0.3--0.4.}
\titlerunning{Initial mass function of planetesimals formed by the streaming instability}
\authorrunning{Sch\"{a}fer et al.}
\maketitle

\section{Introduction}
One of the greatest problems in the theory of planet formation is to explain how millimeter- or centimeter-sized solid particles -- in the following referred to as pebbles -- grow to kilometer-sized planetesimals. Micron-sized dust grains can grow to pebble sizes by coagulation, but larger particles bounce or fragment under mutual collisions \citep{Guettler2010, Zsom2010, Birnstiel2011}. Growth might continue despite this so-called bouncing barrier for very porous ice particles \citep{Wada2008, Wada2009} or by mass transfer in high-speed collisions \citep{Wurm2005, Windmark2012}. 

A fundamental problem of planetesimal formation is the time constraint inflicted by the radial drift of solid particles, a problem that persists even under the assumption of perfect sticking. The orbital velocity of the gas in a protoplanetary disk is sub-Keplerian because the gas is supported against the gravity of the central star by a radial pressure gradient. The gas exerts a drag force on the particles in the disk, whose orbital speed would be equal to the Keplerian speed if the gas was not present, causing them to lose angular momentum and drift radially towards the star. The drift velocity depends on the size of the particles, but is in general highest for meter-sized particles in the inner regions of the disk. A particle with a size of 1~m, initially orbiting at a distance of 1~au from the star, drifts towards the star and sublimates in less than 100~years \citep{Adachi1976, Weidenschilling1977, Brauer2007}. Hence, there needs to be some mechanism assisting the growth of pebbles into planetesimals, which are sufficiently large for the effect of the drag force exerted on them by the gas to be negligible, on a timescale shorter than the radial drift timescale.

The streaming instability provides a mechanism to concentrate solid materials and form planetesimals despite the poor sticking efficiency of the particles and their radial drift. It was discovered analytically by \citet{Youdin2005} and confirmed numerically by \citet{Youdin2007}, \citet{Johansen2007a}, and \citet{Bai2010a}. The radial drift speed of solid particles decreases with increasing solid-to-gas density ratio because of the drag force exerted by the particles on the gas. A locally enhanced solid-to-gas ratio causes the local orbital velocity of the gas to be closer to Keplerian, and thus a reduction of the local drift speed of the particles. Hence, clusters of particles drift more slowly than isolated particles, and downstream clusters can accumulate upstream isolated particles, further reducing the drift speed of the clusters. Owing to this positive feedback loop, particles can be concentrated into filaments reaching maximum densities of up to several thousand times the local gas density \citep{Bai2010a, Johansen2012, Yang2014, Johansen2015}, sufficient to undergo gravitational collapse and form planetesimals \citep{Johansen2007b, Simon2016}. For this strong clustering of particles to occur, the solid-to-gas column density ratio needs to exceed a critical value \citep{Johansen2009b, Bai2010b}, which depends on the radial pressure gradient supporting the gas \citep{Bai2010c} and the particle size \citep{Carrera2015, Yang2016}.

Although several studies have shown that the streaming instability can lead to the formation of planetesimals, their birth mass distribution has not been comprehensively investigated. However, the initial mass function of planetesimals is essential for the study of the formation of planetary systems because it determines the initial conditions for the evolution of the bodies that planetesimals evolve into, including planets, asteroids, and Kuiper belt objects. The asteroids in the asteroid belt provide a natural sample distribution that can be fitted with a broken power law. \citet{Bottke2005} argue that the current size distribution of asteroids larger than 120 km in diameter represents the birth size distribution of the planetesimals that formed in the asteroid belt (but have been strongly depleted by resonances with Jupiter, independent of their sizes), while smaller asteroids are largely fragments of collisions between the larger ones.

Both \citet{Johansen2015} and \citet{Simon2016} performed numerical simulations of planetesimal formation by the streaming instability and find that the differential distribution of the planetesimal birth masses is well-fitted with a power law with an exponent of about~$-1.6$, albeit with the difference that, while the former observe an exponential tapering of the power-law distribution that constitutes the physical upper mass cutoff, the latter do not include such a tapering in their fits. In this paper, we compare power-law fits with and without exponential cutoff to evaluate how well the high-mass end of the initial mass function is described by an exponential cutoff.

\citet{Johansen2015} find the shape of the initial mass function to be relatively independent of the resolution of the simulations and the solid particle column density. They show that a higher resolution leads to the formation of planetesimals with a wider range of sizes, between 30~km and 120~km in radius in their simulation with the highest resolution because the size of the smallest planetesimal declines with increasing resolution. On the other hand, they observe the size of the largest planetesimal to mainly depend on the particle column density, with smaller column densities yielding smaller sizes. \citet{Simon2016} also studied the dependence of the shape of the birth mass distribution on the resolution of the simulations and obtain the same result. They further find the shape of the distribution to be largely independent of the strength of the self-gravity and the simulation time at which it is initiated, although the masses of the planetesimals are shifted to higher values with the increasing strength of self-gravity. The planetesimals that formed in their simulations typically range in radius from 50~km to a few hundred kilometers.

It remains unclear if the planetesimal initial mass function depends on the dimensions of the simulation box. Both \citet{Johansen2015} and \citet{Simon2016} employed only one box size of 0.2~gas scale heights in the radial, azimuthal, and vertical directions. However, \citet{Yang2014} find that, while in the simulations with this box size, the solid particles are concentrated by the streaming instability into only one axisymmetric filament, multiple of these filaments form in simulations with larger box dimensions. This raises the question of whether the mass budget of planetesimal formation, and thus the shape of the initial mass function, is different when not only one filament is observed. In this paper, we study simulations with three different box sizes, the smallest of which is equal to that employed by \citet{Johansen2015} and \citet{Simon2016}, while the others are two and four times larger, respectively, in the radial and azimuthal directions, which permits investigating planetesimal formation in several filaments. Furthermore, in simulations with larger box sizes, more planetesimals emerge, yielding better statistics in particular for the determination of the initial mass function.

The paper is structured as follows: In Sect. 2, the simulation setup, i.e.\ the initial conditions and the parameters that govern the evolution of the simulations are described. In Sect. 3, we present our results regarding the formation of planetesimals by the streaming instability and their radial migration. Further, we comment on the issue of permitting the mutual accretion of sink particles, which we use to model planetesimals. In Sect. 4, we discuss whether the planetesimal birth mass distribution is exponentially tapered and how its shape depends on the dimensions of the simulation box as well as the resolution. We conclude in Sect. 5.

\section{Simulation setup}
\begin{table}[t]
\caption{Simulation specifications}
\label{specifications}
\resizebox{\hsize}{!}{
\begin{tabular}{lccc}
\hline
\hline
Name&$L_x~[H_{\rm{g}}]\times L_y~[H_{\rm{g}}]\times L_z~[H_{\rm{g}}]$\tablefootmark{a}&Resolution [$H_{\rm{g}}^{-1}$]&$N_x\times N_y\times N_z$\tablefootmark{b}\\
\hline
\textit{run\_0.2\_320}&$0.2\times 0.2\times 0.2$&320&$64\times 64\times 64$\\
\hline
\textit{run\_0.4\_320}&$0.4\times 0.4\times 0.2$&320&$128\times 128\times 64$\\
\hline
\textit{run\_0.8\_320}&$0.8\times 0.8\times 0.2$&320&$256\times 256\times 64$\\
\hline
\textit{run\_0.2\_640}&$0.2\times 0.2\times 0.2$&640&$128\times 128\times 128$\\
\hline
\textit{run\_0.4\_640}&$0.4\times 0.4\times 0.2$&640&$256\times 256\times 128$\\
\end{tabular}
}
\tablefoot{
\tablefoottext{a}{Box dimensions in the $x$-,$y$-, and $z$-directions, where $H_{\rm{g}}$ is the gas scale height.}
\tablefoottext{b}{Number of grid cells in the $x$-,$y$-, and $z$-directions.}
}
\end{table}

We conduct three-dimensional computer simulations with the Pencil Code\footnote{\url{http://pencil-code.nordita.org/}}, a hybrid code for gas, for which the magnetohydrodynamic equations are solved on a fixed grid, with Lagrangian particles representing solid bodies. The code employs sixth-order finite differences in space and third-order Runge-Kutta steps in time.

We use the shearing box approximation \citep{Goldreich1965}, i.e.\ we assume that the size of the simulation box is small compared to the distance to the central star of the protoplanetary disk. Hence, the curvature of the disk is neglected and the stellar gravity is linearized. The rectangular simulation box is aligned such that the $x$-, $y$-, and $z$-directions correspond to the radial, azimuthal, and vertical directions, respectively, and co-rotates with the Keplerian velocity at its origin. For both gas and particles, sheared periodic boundary conditions are employed at the radial and azimuthal boundaries and periodic boundary conditions at the vertical boundaries \citep{Hawley1995, Brandenburg1995, Youdin2007, Johansen2009a}.

In total, we perform five simulations with three different simulation box dimensions and two different resolutions, as listed in Table~\ref{specifications}. The two smaller boxes have a size of 0.2 and 0.4~gas scale heights, respectively, in the radial and azimuthal directions with a resolution of either 320 or 640~grid cells per scale height, while the largest box has a radial and azimuthal size of 0.8~scale heights with a resolution of 320~grid cells per scale height. All simulation boxes have a vertical size of 0.2~scale heights. The names of the simulations are composed of the radial and azimuthal dimension as the first number and the resolution as the second number.

\subsection{Gas}
The simulation box is filled with an isothermal, non-magnetized gas with an isothermal equation of state \mbox{$p_{\rm{g}}=c_{\rm{s}}^2\rho_{\rm{g}}$}, where $p_{\rm{g}}$ and $\rho_{\rm{g}}$ are the pressure and density, respectively, and $c_{\rm{s}}$ is the (constant) sound speed. While the gas density is initially constant in the radial and azimuthal direction, it is stratified in the vertical direction because we take into consideration the vertical gravity of the central star, which causes both gas and solid particles to sediment to the mid-plane at \mbox{$z=0$}. This background density stratification is determined by the equilibrium between vertical gravity and vertical pressure gradient, and is given by
\begin{equation}
\rho_{\rm{g}}(z)=\rho_{\rm{g,0}}\exp\left(-\frac{z^2}{2H_{\rm{g}}^2}\right),
\label{rho_g}
\end{equation}
where \mbox{$H_{\rm{g}}=c_{\rm{s}}/\Omega_{\rm{K}}$} is the gas scale height, \mbox{$\Omega_{\rm{K}}=2\pi/P_{\rm{K}}$} the Keplerian orbital frequency, and  $P_{\rm{K}}$ the Keplerian orbital period. Here and in the following, the subscript zero refers to the mid-plane. As formulated in \citet{Yang2014}, we subtract the background density stratification from the equations of the motion for the gas to numerically balance this equilibrium state down to machine precision.

Since the gas density is initially radially constant, there is no radial pressure gradient to support the gas and cause it to orbit with sub-Keplerian speed. Hence, a background pressure gradient set by the dimensionless parameter
\begin{equation}
\Pi=-\frac{1}{2}\frac{H_{\rm{g}}}{R}\frac{\partial\ln (p_{\rm{g,0}})}{\partial\ln (R)}=0.05,
\end{equation}
where $R$ is the orbital distance, is imposed. We refer to this parameter as $\Pi$ adopting the notation by \citet{Bai2010b}. The resulting sub-Keplerian orbital velocity of the gas is given by \mbox{$v_{\rm{g}}=v_{\rm{K}}-\Delta v$}, where \mbox{$v_{\rm{K}}=\Omega_{\rm{K}}R$} is the Keplerian orbital velocity and \mbox{$\Delta v=\Pi c_{\rm{s}}=0.05c_{\rm{s}}$}, which is a representative value at orbital distances of the order of 1~au in a typical protoplanetary disk \citep{Hayashi1981, Bai2010b, Bitsch2015}.

\subsection{Particles}
Two types of Lagrangian particles are employed in the simulations: super-particles and sink particles to model pebbles and planetesimals, respectively. To achieve a good load balancing among the processors, we use the particle block domain decomposition algorithm implemented by \citet{Johansen2011}. For the calculation of the mutual drag forces between the gas and the super-particles and the mutual gravitational forces between the super-particles and the sink particles, we apply the triangular shaped cloud scheme to map the particle masses and velocities onto the grid, and similarly interpolate the back-reaction drag forces and the self-gravitational forces onto the particles \citep{Hockney1981, Youdin2007, Johansen2007b}. 

The gravitational potential of the super-particles as well as the sink particles is computed by solving Poisson's equation using the fast Fourier transform algorithm \citep{Gammie2001}, which entails gravitational softening. The softening length is of the order of the grid cell edge length. Even though the gravitational potential of the particles is vertically periodic, the particles are concentrated in a thin layer around the mid-plane such that their dynamics are not affected by the periodic potential away from the mid-plane. 

We neglect the contribution of the gas to the gravitational potential under the assumption that the gas component of the protoplanetary disk is not in the gravitationally unstable regime. Indeed, in our simulations the gas density deviates by at most 2\% from the mid-plane density~$\rho_{\rm{g,0}}$, the gas density perturbations are thus very small compared to the super-particle densities in the filaments when planetesimals form, which are of the order of the Roche density (see below).

The sedimentation of the super-particles to the mid-plane that is due to the vertical stellar gravity induces turbulence as a result of either the streaming instability or the Kelvin-Helmholtz instability, which are both caused by the mutual drag forces between the gas and the super-particles~\citep{Bai2010b}. This turbulence stirs up the super-particles, and hence counteracts the sedimentation. To give sedimentation and turbulence time to attain an equilibrium, in our simulations self-gravity is not introduced until \mbox{$t=25~P_{\rm{K}}$}. Its strength is then gradually increased from zero over 10~$P_{\rm{K}}$ until it reaches its final value at \mbox{$t=35~P_{\rm{K}}$} since initiating self-gravity instantaneously with full strength could cause significant impulses on the particles. While we initiate self-gravity at a simulation time at which the super-particles have already formed filaments, \citet{Johansen2015} introduced self-gravity with full strength at the start of their simulations and observe qualitatively the same planetesimal birth mass distribution as we do.

We achieve this gradual initiation of the self-gravity by substituting
\begin{equation}
\Gamma=
\begin{cases}
0&t\leq 25~P_{\rm{K}},\\
\frac{1}{2}\gamma\left(1-\cos\left[\frac{\pi(t-25~P_{\rm{K}})}{10~P_{\rm{K}}}\right]\right)&25~P_{\rm{K}}<t<35~P_{\rm{K}},\\
\gamma&t\geq 35~P_{\rm{K}},\\
\end{cases}
\end{equation}
where the dimensionless self-gravity parameter
\begin{equation}
\gamma=\frac{4\pi G\rho_{\rm{g,0}}}{\Omega_{\rm{K}}^2}
\label{gamma}
\end{equation}
and $G$ is the gravitational constant, into the right-hand side of Poisson's equation. We choose \mbox{$G=P_{\rm{K}}^{-2}\rho_{\rm{g,0}}^{-1}$}, and thus \mbox{$\gamma=1/\pi=0.318$}. We note that \citet{Simon2016} find the shape of the initial mass function of the planetesimals formed in their simulations of the streaming instability to be relatively independent of both the simulation time at which self-gravity is introduced and the strength of the self-gravity. The Roche density depends on the self-gravity parameter $\gamma$ and is given by
\begin{equation}
\rho_{\rm{R}}=\frac{9\Omega_{\rm{K}}^2}{4\pi G}=\frac{9\rho_{\rm{g,0}}}{\gamma}=28.3\rho_{\rm{g,0}}.
\label{rho_R}
\end{equation}

Each super-particle represents a large number of equally sized pebbles because it is computationally infeasible to simulate the pebbles individually. While the mass of a super-particle is equal to the total mass of the pebbles it models, its friction time is the same as that of an individual constituent pebble. 

The mass of the super-particles is determined by the initial solid-to-gas column density ratio and their initial number. We set the solid-to-gas ratio 
\begin{equation}
Z=\frac{\Sigma_{\rm{p,init}}}{\Sigma_{\rm{g,init}}},
\end{equation}
where $\Sigma_{\rm{p,init}}$ and 
\begin{equation}
\Sigma_{\rm{g,init}}=\sqrt{2\pi}H_{\rm{g}}\rho_{\rm{g,0}}
\end{equation}
are the initial column densities of the super-particles and the gas, respectively, to $Z=0.02$. This value corresponds to the critical solid-to-gas ratio necessary for strong clustering of pebbles due to the streaming instability to occur~\citep{Johansen2009b, Bai2010b, Bai2010c, Carrera2015}, and is slightly higher than the solar metallicity. The initial number of super-particles is set equal to the total number of grid cells. The super-particles are randomly distributed among the entire simulation box to seed the streaming instability.

The Stokes number of the super-particles \mbox{$\tau_{\rm{f}}=\Omega_{\rm{K}}t_{\rm{f}}$}, where $t_{\rm{f}}$ is the friction time, is set to \mbox{$\tau_{\rm{f}}=\pi/10=0.314$}, which, at an orbital distance of 2.5~au in the Minimum Mass Solar Nebula (MMSN), corresponds to a size of approximately 25~cm \citep{Bai2010b, Johansen2015}. While we employ only one fixed particle size, \citet{Bai2010b} performed simulations with particles of a range of sizes. They find that the particle-gas dynamics are dominated by the most massive particles, and that the critical solid-to-gas ratio required for strong particle clustering owing to the streaming instability is determined by the total mass of all particles.

After self-gravity has attained its full strength at \mbox{$t=35~P_{\rm{K}}$}, every super-particle comprised in a cluster whose super-particle density $\rho_{\rm{p}}$ exceeds a threshold value $\rho_{\rm{p,thres}}$ is replaced by a sink particle. This sink particle creation threshold is set to \mbox{$\rho_{\rm{p,thres}}=200~\rho_{\rm{g,0}}$}, i.e.\ about seven times the Roche density (see Eq.~\ref{rho_R}). We note that \citet{Johansen2015} compared simulations similar to ours with three different threshold values and find the masses of the sink particles emerging in their simulations to be largely independent of the threshold above a value of five times the Roche density. 

The simulation time at which the formation of sink particles is introduced is arbitrary since both the gravitationally bound super-particle clusters that exist beforehand and the sink particles that emerge from them afterwards represent planetesimals. Owing to the limited resolution of the gravitational forces the behavior of many super-particles inside one grid cell is comparable to that of a few sink particles. Nevertheless, the computational expense of the simulations is lowered substantially by the introduction of sink particles.

Super-particles within the accretion radius of a sink particle, which is set equal to one grid cell edge length, are accreted by it, i.e.\ the super-particle mass and momentum are added to the sink particle mass and momentum, respectively, and the super-particle is removed. This accretion, however, might in parts be artificial because the physical accretion radius, i.e.\ the Bondi radius for pebble accretion \citep{Ormel2010, Lambrechts2012}, could be smaller than the simulated accretion radius, especially in the case of less massive sink particles. On the other hand, since the mutual gravitational forces between the super-particles and the sink particles within one grid cell can only be computed inaccurately, the chosen accretion radius corresponds to the highest accuracy our numerical simulations can offer.

We further permit sink particles to accrete one another. This accretion is handled analogously to the super-particle accretion, and might as well be partially artificial. Nevertheless, it is required because in a super-particle cluster exceeding the sink particle creation threshold, all super-particles are replaced by sink particles. That is, the gravitational collapse of a super-particle cluster results in the creation of a cluster of sink particles, of which only one should persist to represent one new-born planetesimal. See Sect. 3.2 for further discussion of this topic. 

\subsection{Units and scaling relations}
We report our results using the Keplerian orbital period at the origin of the simulation box $P_{\rm{K}}$, the gas scale height $H_{\rm{g}}$, and the mid-plane gas density $\rho_{\rm{g,0}}$ as the units of time, length, and density, respectively. We note that a shearing box freely scales with these units until self-gravity is initiated at \mbox{$t=25~P_{\rm{K}}$}. Afterwards,
\begin{equation}
\rho_{\rm{g,0}}=\frac{\gamma\Omega_{\rm{K}}^2}{4\pi G}=\frac{\pi\gamma}{GP_{\rm{K}}^2}
\end{equation}
(see Eq.~(\ref{gamma})), and hence the unit of mass \mbox{$[M]=H_{\rm{g}}^3\rho_{\rm{g,0}}=\pi\gamma~G^{-1}~H_{\rm{g}}^3P_{\rm{K}}^{-2}$}. 

In the following, relations for the scaling of relevant quantities and units with the orbital distance $R$, the temperature $T$, and the mass of the central star $M_{\rm{S}}$ are given. A mean atomic weight of \mbox{$\mu=2.33$} is used and the scaling relations for $\rho_{\rm{g,0}}$ and $[M]$ are calculated applying those for $P_{\rm{K}}$ and $H_{\rm{g}}$: 
\begin{equation}
P_{\rm{K}}=4.0~\left(\frac{R}{2.5~\rm{au}}\right)^{3/2}~\left(\frac{M_{\rm{S}}}{1~\rm{M_{\sun}}}\right)^{-1/2}~\rm{yr}
\end{equation}
\begin{equation}
\Omega_{\rm{K}}=1.6~\left(\frac{R}{2.5~\rm{au}}\right)^{-3/2}~\left(\frac{M_{\rm{S}}}{1~\rm{M_{\sun}}}\right)^{1/2}~\rm{yr}^{-1}
\end{equation}
\begin{equation}
c_{\rm{S}}=0.80~\left(\frac{T}{180~\rm{K}}\right)^{1/2}~\rm{km}~\rm{s}^{-1}
\end{equation}
\begin{equation}
H_{\rm{g}}=0.11~\left(\frac{R}{2.5~\rm{au}}\right)^{3/2}~\left(\frac{T}{180~\rm{K}}\right)^{1/2}~\left(\frac{M_{\rm{S}}}{1~\rm{M_{\sun}}}\right)^{-1/2}~\rm{au}
\label{H_g}
\end{equation}
\begin{equation}
\rho_{\rm{g,0}}=9.4\times10^{-10}~\left(\frac{\gamma}{\pi^{-1}}\right)~\left(\frac{R}{2.5~\rm{au}}\right)^{-3}~\left(\frac{M_{\rm{S}}}{1~\rm{M_{\sun}}}\right)~\rm{g}~\rm{cm}^{-3}
\end{equation}
\begin{equation}
\begin{split}
[M]=4.2\times10^{27}~\left(\frac{\gamma}{\pi^{-1}}\right)~\left(\frac{R}{2.5~\rm{au}}\right)^{3/2}~\left(\frac{T}{180~\rm{K}}\right)^{3/2}\\
\times\left(\frac{M_{\rm{S}}}{1~\rm{M_{\sun}}}\right)^{-1/2}~\rm{g}
\label{[M]}
\end{split}
\end{equation}

Hereafter, we use the above scaling relations, the properties of the asteroid belt, i.e.\ an orbital distance of \mbox{$R=2.5~\rm{au}$}, a temperature of \mbox{$T=180~\rm{K}$}, and a stellar mass of \mbox{$M_{\rm{S}}=1~\rm{M_{\sun}}$}, and the chosen strength of the self-gravity \mbox{$\gamma=1/\pi$} to convert simulation units into physical units. For example, the mid-plane gas density \mbox{$\rho_{\rm{g,0}}=9.4\times10^{-10}~\rm{g}~\rm{cm}^{-3}$}, which is almost one order of magnitude greater than the corresponding value in the MMSN, \mbox{$\rho_{\rm{g,0}}=1.1\times10^{-10}~\rm{g}~\rm{cm}^{-3}$} \citep{Hayashi1981, Bai2010b}. The streaming instability has been shown to form planetesimals for pebble column densities similar to that in the MMSN \citep{Johansen2015}. Nevertheless, we choose this comparably high gas density and thus high solid density to promote planetesimal formation, enabling us to better constrain the initial mass function of planetesimals.

\section{Evolution of the simulations}
\begin{figure}[H]
\resizebox{\hsize}{!}{\includegraphics[width=0.45\textwidth]{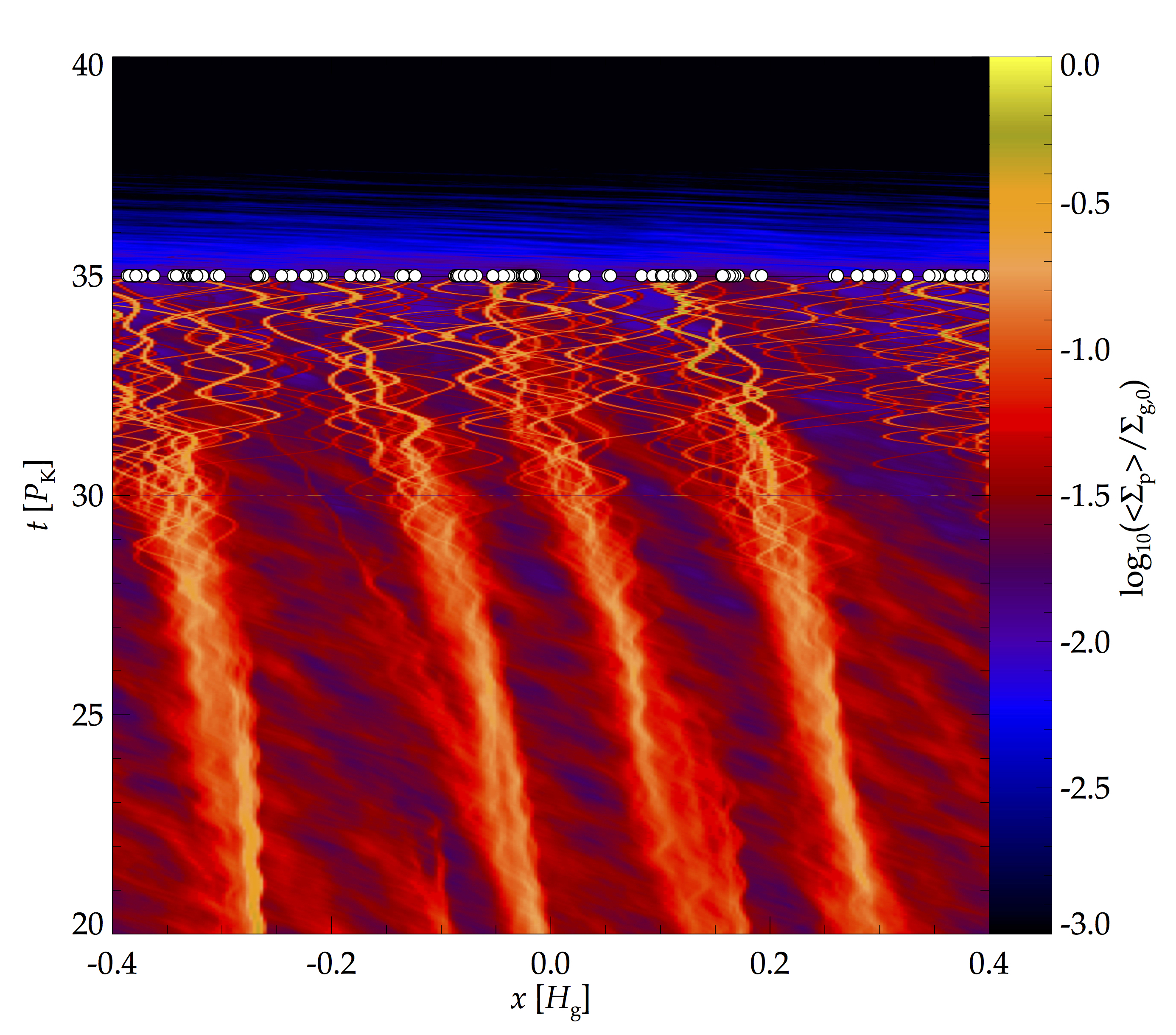}}
\caption{Pebble column density $\Sigma_{\rm{p}}$, integrated over the vertical dimension and averaged over the azimuthal dimension of the simulation box, as a function of radial location $x$ and simulation time $t$ for the simulation with the largest box size, \textit{run\_0.8\_320}. Locations at which sink particles emerge are indicated with white dots. Though every pebble cluster forms in one of the filaments, they migrate up to the distance to one of the adjacent filaments. All sink particles emerge as soon as their formation is initiated at \mbox{$t=35~P_{\rm{K}}$}, and they emerge nearly evenly distributed among the entire radial extent of the box owing to the radial migration of the pebble clusters.}
\label{rhopmx}
\end{figure}

Apart from the use of self-gravity and sink particles to model planetesimals, our simulation setup is identical with that applied by \citet{Yang2014}. Hence, we find our simulations to be consistent with their simulations until \mbox{$t=25~P_{\rm{K}}$}, when self-gravity is introduced (compare the evolution of the pebble density in the simulation time span \mbox{$20~P_{\rm{K}}\leq t\leq 25~P_{\rm{K}}$} shown in Fig.~\ref{rhopmx} with their Fig. 3). We thus only report on the evolution of our simulations between \mbox{$t=25~P_{\rm{K}}$} and the end of the simulations, \mbox{$t=40~P_{\rm{K}}$}.

\subsection{Planetesimal formation and migration}
\begin{figure*}[t]
\centering
\begin{minipage}{0.49\textwidth}
\centering
\includegraphics[width=\textwidth, trim = {1cm 2cm 0cm 2cm}, clip]{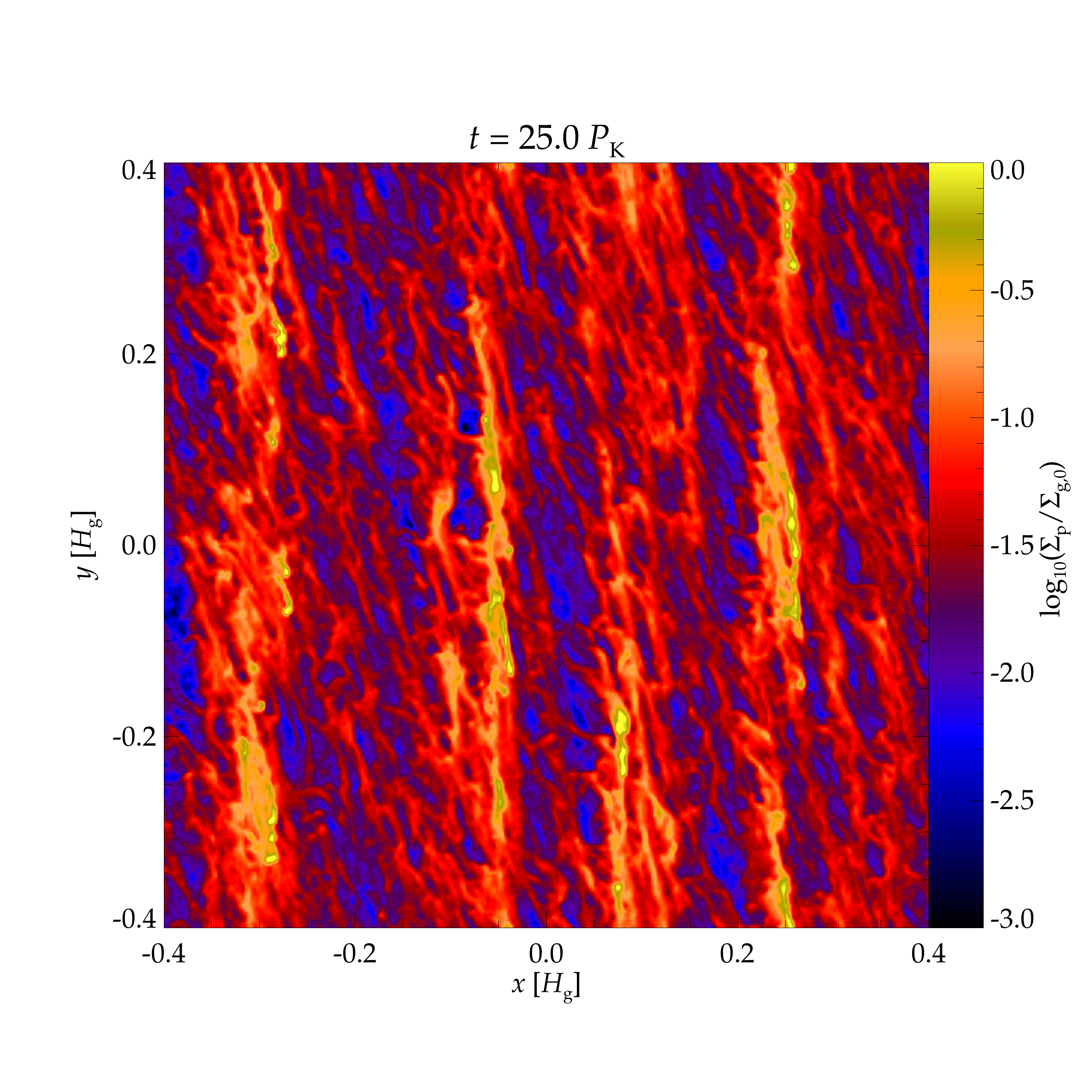}
\includegraphics[width=\textwidth, trim = {1cm 2cm 0cm 2cm}, clip]{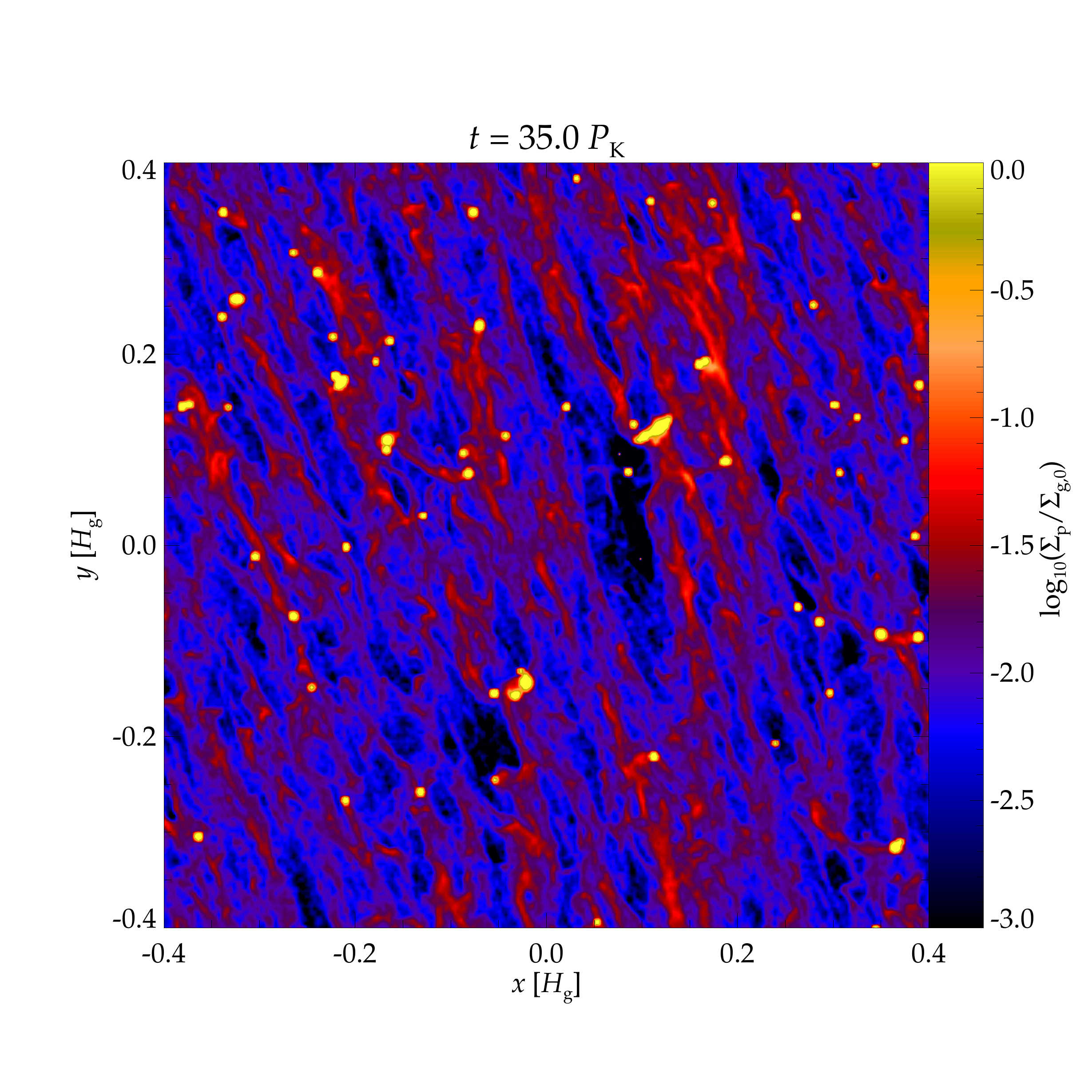}
\end{minipage}
\hfill
\begin{minipage}{0.49\textwidth}
\centering
\includegraphics[width=\textwidth, trim = {1cm 2cm 0cm 2cm}, clip]{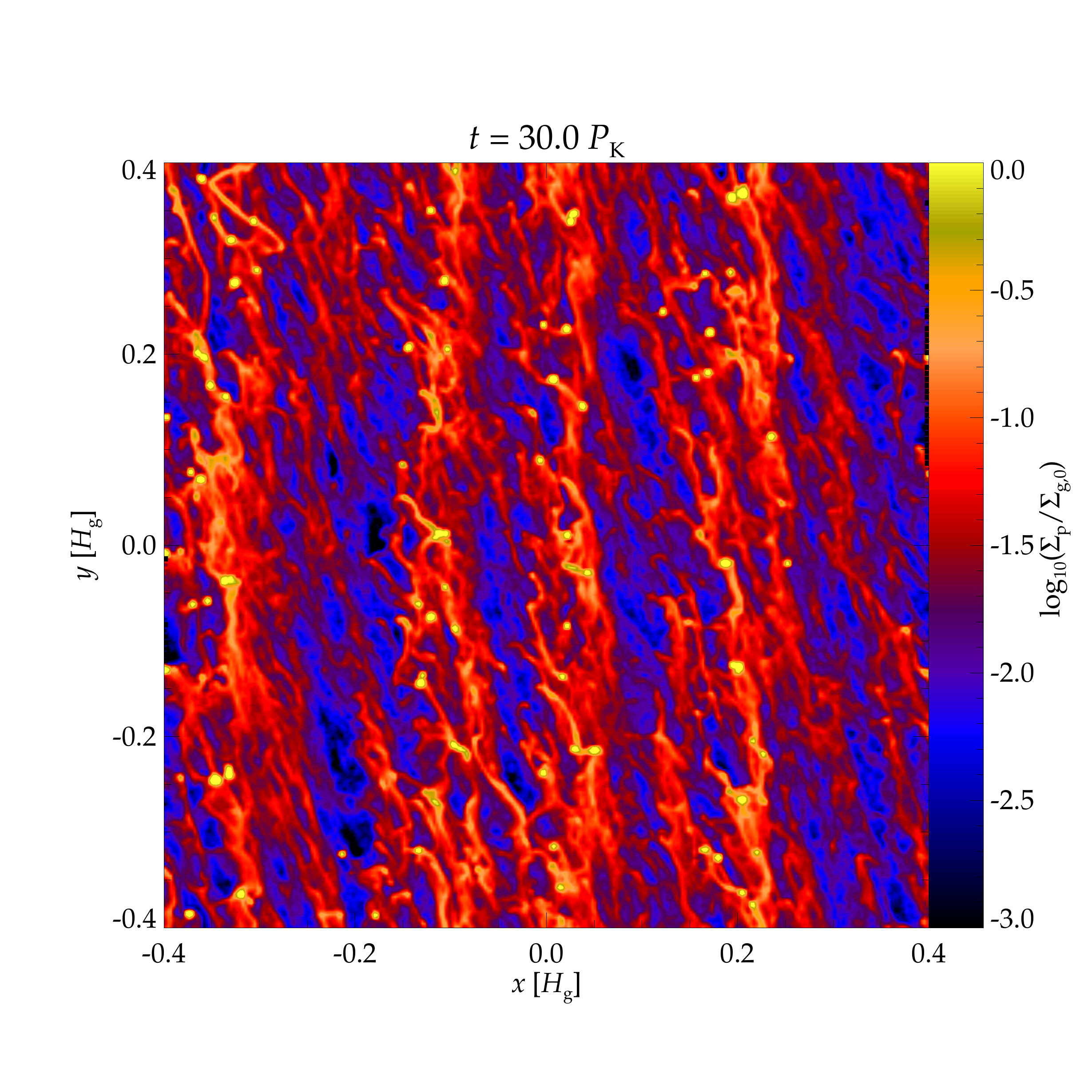}
\includegraphics[width=\textwidth]{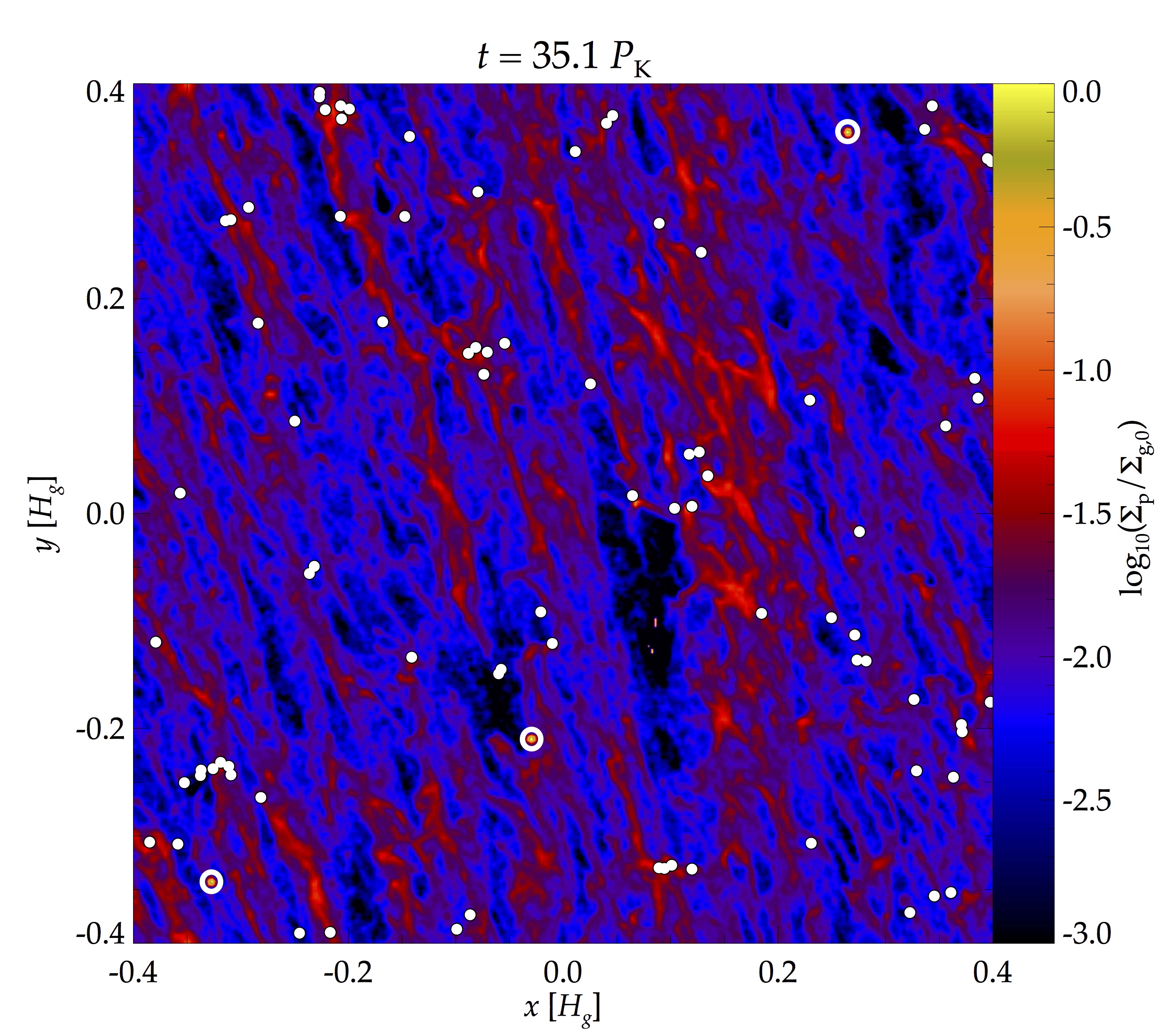}
\end{minipage}
\caption{Pebble column density $\Sigma_{\rm{p}}$, integrated over the vertical dimension of the simulation box, as a function of radial location $x$ and azimuthal location $y$ at four different simulation times \mbox{$t=25~P_{\rm{K}}$} (top left panel), \mbox{$t=30~P_{\rm{K}}$} (top right panel), \mbox{$t=35~P_{\rm{K}}$} (bottom left panel), and \mbox{$t=35.1~P_{\rm{K}}$} (bottom right panel) for the simulation with the largest box size, \textit{run\_0.8\_320}. In the lower right panel, sink particles are plotted as white dots and three pebble clusters are indicated using white circles. After self-gravity has been initiated at \mbox{$t=25~P_{\rm{K}}$}, the pebbles aggregate into clusters and the axisymmetric filaments disperse (upper panels). When the self-gravity attains its full strength at \mbox{$t=35~P_{\rm{K}}$}, most pebbles are concentrated in clusters and the filaments are no longer visible (lower left panel). At this point, we introduce the formation of sink particles, and the pebbles comprised in all but the three encircled clusters are replaced by sink particles. The sink particles emerging from the same pebble cluster undergo a merging process until only one of them remains. This process is largely completed at \mbox{$t=35.1~P_{\rm{K}}$} (lower right panel).}
\label{rhopmxy}
\end{figure*}

If self-gravity is not taken into account the streaming instability concentrates pebbles into axisymmetric filaments that are elongated in the azimuthal direction \citep{Johansen2007a, Johansen2009b, Bai2010b, Yang2014}. In Figs.~\ref{rhopmx} and \ref{rhopmxy}, we show how self-gravity causes these filaments to fragment into pebble clusters that undergo gravitational collapse and form planetesimals. In the first $5~P_{\rm{K}}$ after self-gravity has been initiated, the filaments begin to disperse because the pebbles accumulate into clusters owing to their mutual gravitational attraction (upper panels of Fig.~\ref{rhopmxy}). At \mbox{$t=35~P_{\rm{K}}$}, when the self-gravity reaches its full strength, these clusters contain most pebbles available in the simulation, and the filaments are no longer observable (lower left panel). 

At this point, we commence the formation of sink particles. Almost all pebble clusters exceed the sink particle creation threshold, consequently the pebbles in each of these clusters are replaced by sink particles which merge into one massive sink particle that represents the gravitationally collapsed cluster. At \mbox{$t=35.1~P_{\rm{K}}$}, this sink particle merging process is for the most part completed (lower right panel). However, a few low-mass pebble clusters do not turn into sink particles. All sink particles emerge instantly at \mbox{$t=35~P_{\rm{K}}$}, as can be seen from Fig.~\ref{rhopmx}, but a couple of clusters remain at \mbox{$t=35.1~P_{\rm{K}}$} (three such clusters can be spotted in the lower right panel of Fig.~\ref{rhopmxy}). Although they are not sufficiently dense to exceed the sink particle creation threshold, these clusters probably represent gravitationally bound planetesimals with low masses.

\begin{figure}[t]
\resizebox{\hsize}{!}{\includegraphics[width=0.45\textwidth, trim = {0cm 1cm 1cm 2cm}, clip]{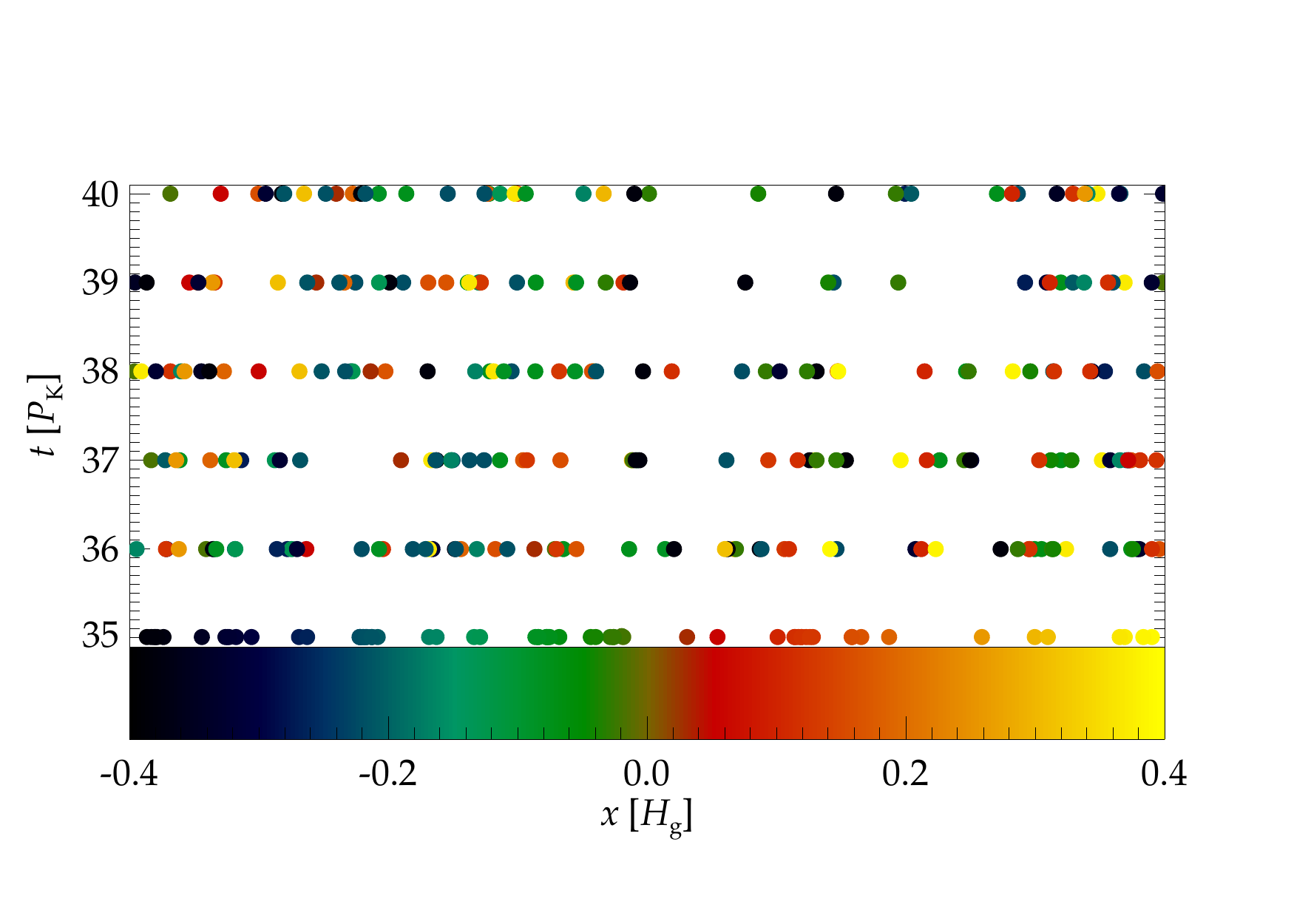}}
\caption{Radial locations $x$ of sink particles which are not accreted before \mbox{$t=36~P_{\rm{K}}$}, color-coded according to the locations at which they emerge, as functions of the simulation time $t$ for the simulation with the largest box size, \textit{run\_0.8\_320}. On average, the motions of the sink particles span more than half of the radial extent of the simulation box, a couple of them even migrate through the entire radial dimension of the box.}
\label{radial locations}
\end{figure}

We observe that the planetesimals on average move through more than half of the radial extent of the simulation boxes. Figure~\ref{rhopmx} shows that each pebble cluster forms in one of the filaments, but that they migrate in the radial direction, some of them only marginally, others the entire distance to one of the adjacent filaments. As a result, the sink particles emerge almost evenly distributed among the whole radial dimension of the box. From Fig.~\ref{radial locations}, it can be seen that the sink particles continue this radial migration, they on average pass through over half of, a few of them even through the whole radial extent of the box. The mean standard deviation of the radial displacement of the sink particles from the locations at which they emerge, averaged over and weighted by the lifetime of every sink particle and then averaged over all sink particles in a simulation, amounts to between 26\% and 36\% of the radial box size in each of the five simulations. We note that the extent of the migration in the radial direction increases with the box size without converging for the box sizes we consider. The radial motions result from the mutual gravitational scattering of sink particles that closely pass by each other. It remains to be investigated whether planetesimals are composed of not only pebbles from the filament they form in, but also of an appreciable amount of pebbles from filaments they migrate to.

\subsection{Mutual sink particle accretion}
For the sink particles to realistically represent new-born planetesimals, only one of them should be allowed to form from every pebble cluster that undergoes gravitational collapse. However, this requires a precise determination of the extent of each cluster, which entails several issues, for instance the treatment of overlapping clusters. Therefore, we replace every super-particle that is part of a cluster that exceeds the sink particle creation threshold by a sink particle and allow the sink particles to accrete one another until only one of them remains. We observe that in our five simulations, on average 81\% of all accreted sink particles are accreted within $0.1~P_{\rm{K}}$ after their formation, i.e.\ until \mbox{$t=35.1~P_{\rm{K}}$}, because in all simulations all sink particles emerge togather at \mbox{$t=35~P_{\rm{K}}$}. Therefore, the merging process of sink particles that emerged from the same pebble cluster is probably largely completed at this point (see the lower right panel of Fig.~\ref{rhopmxy}), and most of the mutual sink particle accretions are a part of this process.

On the other hand, we note that the merging of sink particles afterwards may be artificial. The accuracy of the calculation of the gravitational forces between sink particles is limited by the resolution, hence we cannot determine whether two sink particles that encounter one another inside a grid cell collide or pass by each other. Nevertheless, we find the latter to be more probable. Taking into account gravitational focusing, the maximum impact parameter leading to a collision of two sink particles
\begin{equation}
b_{\rm{max}}=\sqrt{(r_1+r_2)^2+\frac{2G(m_1+m_2)(r_1+r_2)}{\Delta v^2}},
\label{impact parameter}
\end{equation}
where $m_i$ are the masses of the two sink particles, $r_i$ their radii, which are calculated from $m_i$ using a solid body density of~$3~\rm{g}~\rm{cm}^{-3}$, and $\Delta v$ is their relative velocity (see, e.g., \citealt{Armitage2007}). The mean maximum impact parameter $\langle b_{\rm{max}}\rangle $, averaged over all mutual sink particle accretions in all five simulations where the lifetime of the accreted sink particle is greater than 0.1~$P_{\rm{K}}$ and weighted by the lifetimes of the accreted sink particles, amounts to only 3.5\% of the grid cell edge length. Mutual accretions of three or more sink particles at the same simulation time are not included in this statistic because we cannot infer their outcome. We note, though, that the sink particle data are not written out after each simulation time step, but every $0.01~P_{\rm{K}}$. Thus, we can only imprecisely determine the simulation time at which a sink particle merging occurs and the maximum impact parameter for this encounter.

\section{Initial mass function}

We fit the cumulative mass distribution of the sink particles that emerge in our simulations using two functional forms: an integrated power law and a power law with an exponential cutoff. We choose to fit the cumulative mass distributions because, in particular for small numbers of sink particles, they are less affected by noise than the differential mass distributions and can thus be fitted more accurately.

The integrated power law can be expressed as
\begin{equation}
\frac{N_>(M)}{N_{\rm{tot}}}=\frac{1}{1-\alpha}\left[\left(\frac{M_{\rm{max}}}{M_{\rm{pow}}}\right)^{1-\alpha}-\left(\frac{M}{M_{\rm{pow}}}\right)^{1-\alpha}\right],
\label{integrated power law}
\end{equation}
where $N_>$ is the number of sink particles with masses greater than $M$, $N_{\rm{tot}}$ is their total number and $M_{\rm{max}}$ their maximum mass, and $M_{\rm{pow}}$ and $\alpha$ are fitting parameters.
Since the formation of planetesimals by the streaming instability is a stochastic process and the actual $M_{\rm{max}}$ in a simulation might differ significantly from the $M_{\rm{max}}$ of the ensemble-averaged mass distribution of the sink particles for this model, we treat $M_{\rm{max}}$ as a fitting parameter. 

The exponentially tapered power law is given by
\begin{equation}
\frac{N_>(M)}{N_{\rm{tot}}}=\left(\frac{M}{M_{\rm{pow}}}\right)^{-\alpha}~\exp\left[-\left(\frac{M}{M_{\rm{exp}}}\right)^{\beta}\right],
\label{exponentially tapered power law M_pow}
\end{equation}
where $M_{\rm{exp}}$ and $\beta$ are fitting parameters. The condition \mbox{$N_>(M_{\rm{min}})/N_{\rm{tot}}=1$}, where $M_{\rm{min}}$ is the minimum sink particle mass, can be used to eliminate one of the fitting parameters in Eq.~(\ref{exponentially tapered power law M_pow}). We choose to eliminate the characteristic mass of the power-law part, $M_{\rm{pow}}$, which is then given by
\begin{equation}
M_{\rm{pow}}=M_{\rm{min}}~\exp\left[\frac{1}{\alpha}\left(\frac{M_{\rm{min}}}{M_{\rm{exp}}}\right)^{\beta}\right].
\label{M_pow}
\end{equation}
Substituting Eq.~(\ref{M_pow}) into Eq.~(\ref{exponentially tapered power law M_pow}) yields
\begin{equation}
\frac{N_>(M)}{N_{\rm{tot}}}=\left(\frac{M}{M_{\rm{min}}}\right)^{-\alpha}~\exp\left[\left(\frac{M_{\rm{min}}}{M_{\rm{exp}}}\right)^{\beta}-\left(\frac{M}{M_{\rm{exp}}}\right)^{\beta}\right].
\label{exponentially tapered power law M_min}
\end{equation}
We use Eq.~(\ref{exponentially tapered power law M_min}) to fit the sink particle data because we find the resulting fits to be better than the ones we obtain applying Eq.~(\ref{exponentially tapered power law M_pow}). For the same reason as $M_{\rm{max}}$ in Eq.~(\ref{integrated power law}), $M_{\rm{min}}$ is treated as a fitting parameter.

To investigate the dependence of the shape of the initial mass function on the resolution and particularly the dimensions of the simulation box, we determine an individual initial mass function for every simulation. At first, we employ the least-squares method to fit Eqs.~(\ref{integrated power law}) and (\ref{exponentially tapered power law M_min}) to the sink particle data at each Keplerian orbital period between \mbox{$t=36~P_{\rm{K}}$} and the end of the simulations, \mbox{$t=40~P_{\rm{K}}$}. In all five simulations, all sink particles emerge at once at the simulation time at which their formation is initiated, \mbox{$t=35~P_{\rm{K}}$}, but we begin the fitting at \mbox{$t=36~P_{\rm{K}}$} to give the sink particles that emerge from the same pebble cluster time to merge into one. Averaged over this period, we then calculate mean values of the fitting parameters that do not vary significantly with time and are thus probably relatively unaffected by artificial sink particle merging.

\begin{figure*}[t]
\centering
\begin{minipage}{0.49\textwidth}
\centering
\includegraphics[width=\textwidth, trim = {0cm 0cm 0cm 0cm}, clip]{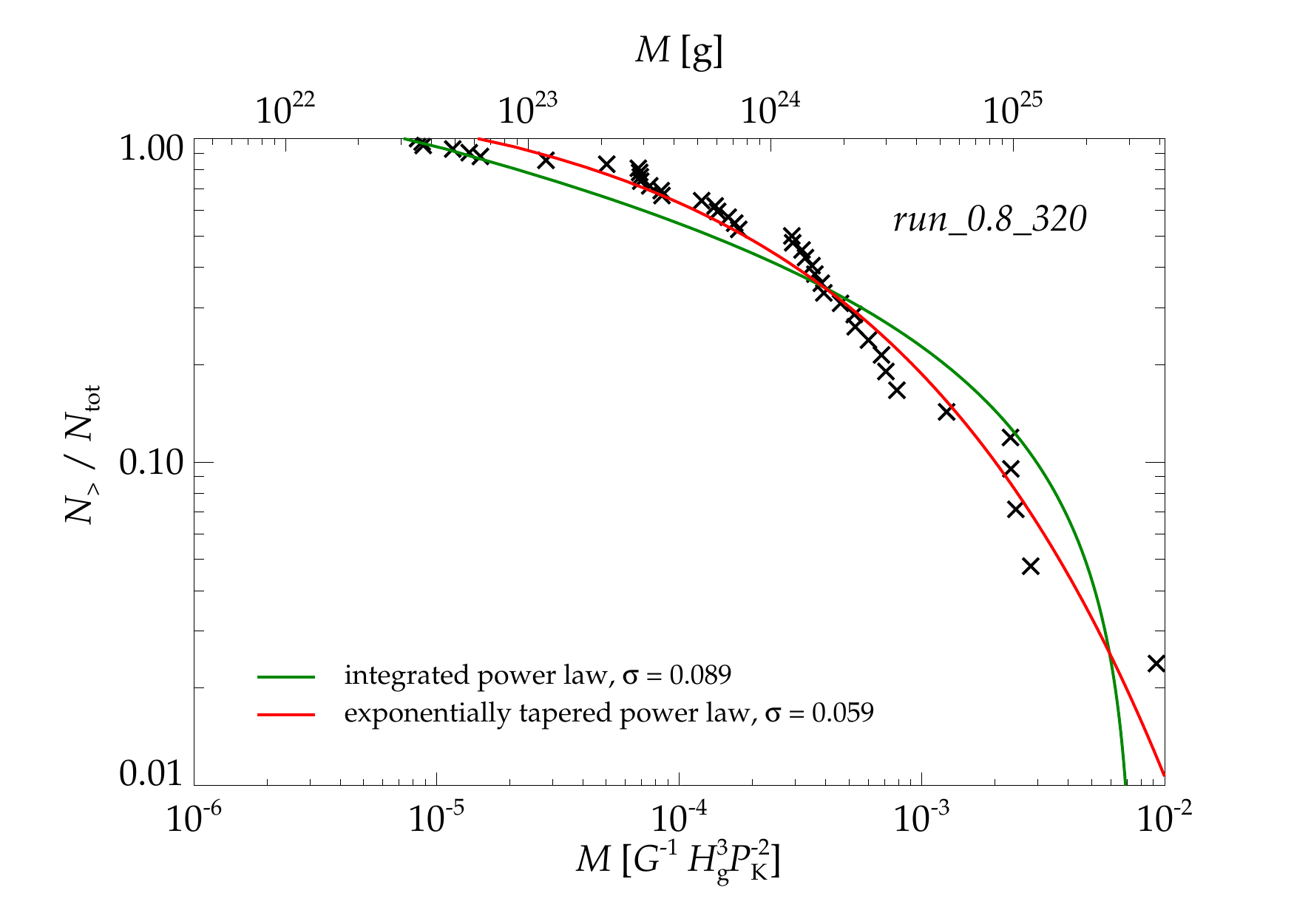}
\end{minipage}
\hfill
\begin{minipage}{0.49\textwidth}
\centering
\includegraphics[width=\textwidth, trim = {0cm 0cm 0cm 0cm}, clip]{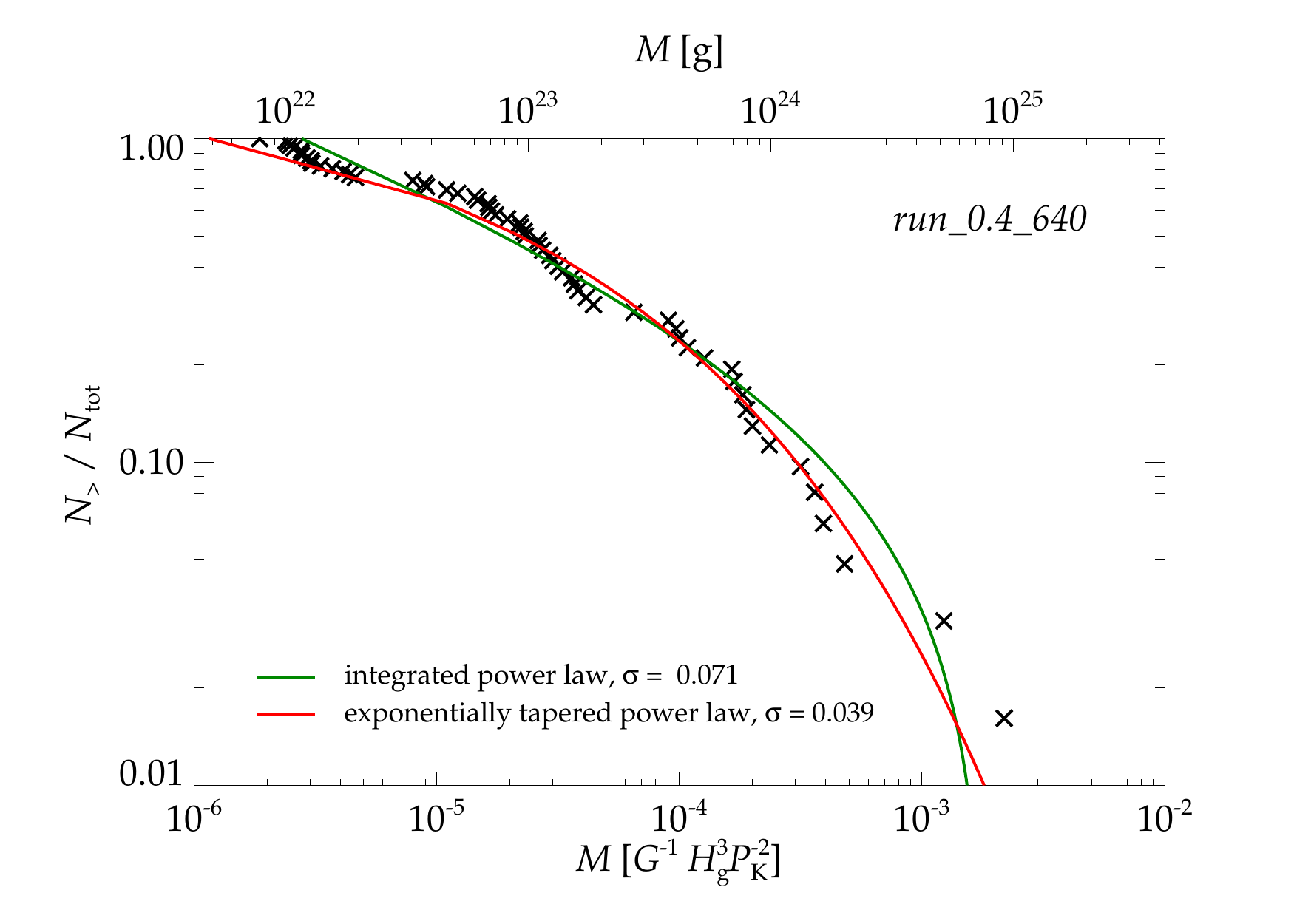}
\end{minipage}
\caption{Cumulative mass distributions of sink particles (black crosses) and fitted integrated power laws (Eq.~\ref{integrated power law}, green lines) and power laws with exponential cutoff (Eq.~\ref{exponentially tapered power law M_min}, red lines) at \mbox{$t=40~P_{\rm{K}}$} for the two simulations with the largest numbers of sink particles: the one with the largest box dimensions, \textit{run\_0.8\_320}, (left panel) and the one with the middle box dimensions and the higher resolution, \textit{run\_0.4\_640} (right panel). Standard deviations of the actual $N_>/N_{\rm{tot}}$ (black crosses) from the fitted $N_>/N_{\rm{tot}}$ are given in the legends. The standard deviations for the integrated power laws are up to twice as large as those for the exponentially tapered power laws. In addition, it can be seen that the shallower exponential cutoffs fit the actual cutoffs more accurately than the steeper cutoffs of the integrated power laws and better replicate the smooth change of the slope of the mass distributions. We thus find the power laws with exponential tapering to represent the mass distributions better than the integrated power laws.}
\label{IMF}
\end{figure*}

In the left panel and the right panel of Fig.~\ref{IMF} we show the sink particle mass distributions as well as the fitted integrated power laws and power laws with exponential tapering for the two simulations with the largest numbers of sink particles, the one with the largest box size, \textit{run\_0.8\_320}, and the one with the middle box size and the higher resolution, \textit{run\_0.4\_640}, respectively, at \mbox{$t=40~P_{\rm{K}}$}. In the legends for both fits, the standard deviation $\sigma$ of the actual \mbox{$N_>(M)/N_{\rm{tot}}$} for the sink particle masses $M$ from the fitted \mbox{$N_>(M)/N_{\rm{tot}}$} is given.
 
We find that the exponential tapered power laws fit the mass distributions better than the integrated power laws. The standard deviations for the power-law fits without exponential cutoff are larger, not only at \mbox{$t=40~P_{\rm{K}}$} as shown in the figure, but also at other simulation times. We further note that the shallower exponential cutoff represents the high-mass end of the distributions better than the steeper cutoff of the integrated power law, and that the exponential tapering better reproduces the smooth change of the slope of the distributions. Hence, we limit our further analysis to the power-law fits with exponential cutoff. 

\subsection{Best-fitting parameters}
\subsection{Power-law fits with and without exponential tapering}

\begin{figure*}[t]
\centering
\begin{minipage}{0.49\textwidth}
\centering
\includegraphics[width=\textwidth, trim = {0cm 0cm 0cm 1cm}, clip]{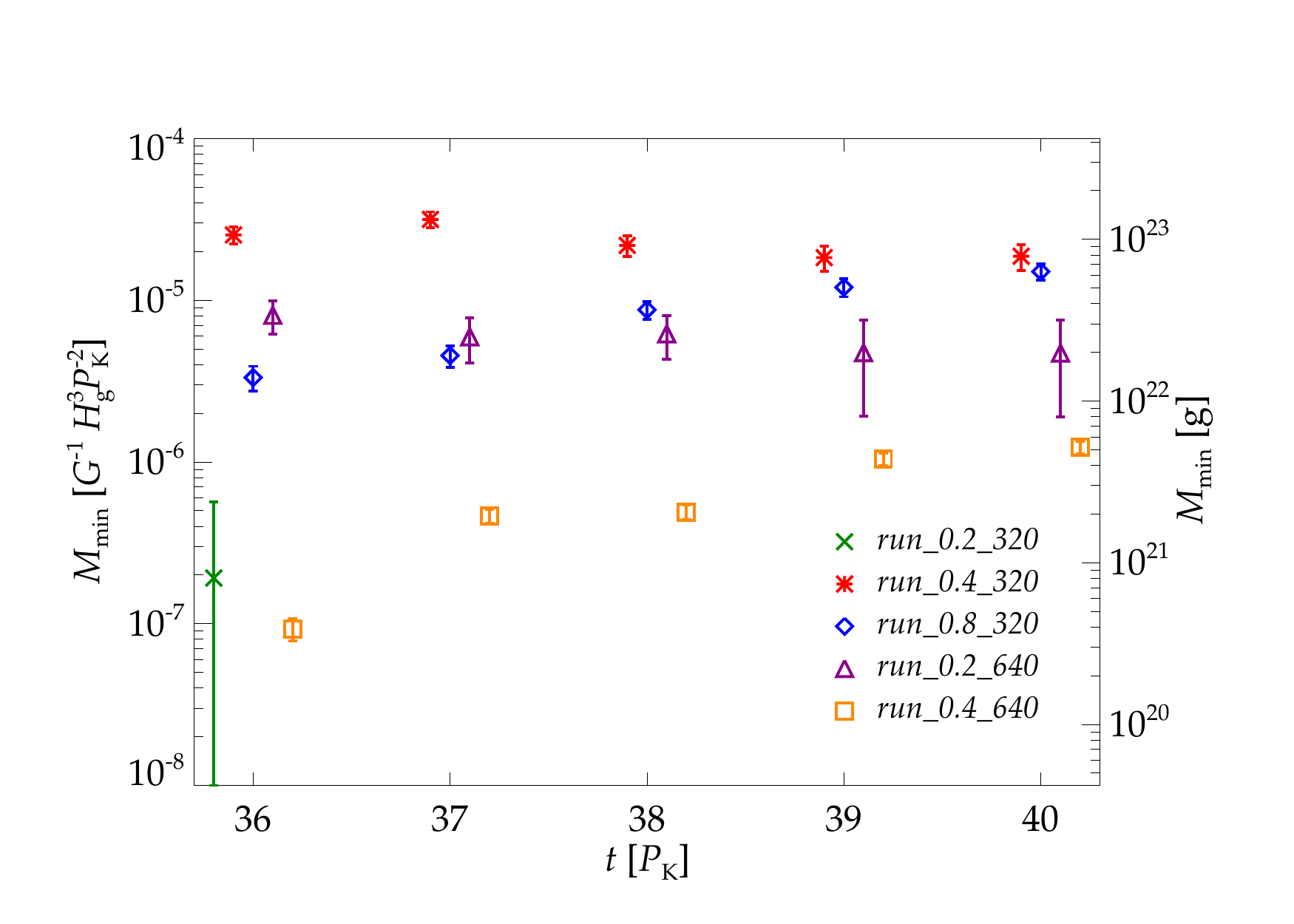}
\includegraphics[width=\textwidth, trim = {0cm 0cm 0cm 1cm}, clip]{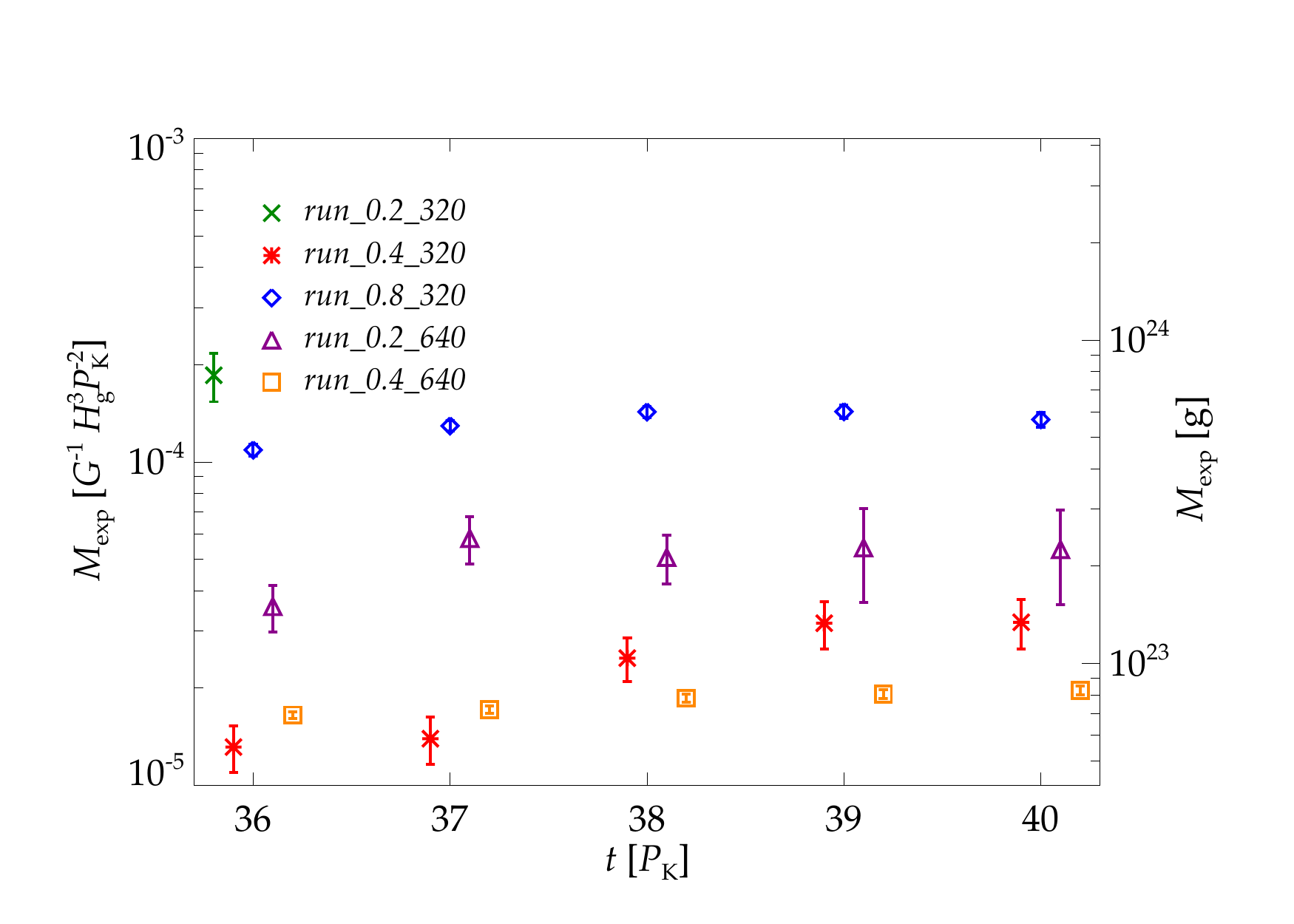}
\end{minipage}
\hfill
\begin{minipage}{0.49\textwidth}
\centering
\includegraphics[width=\textwidth, trim = {0cm 0cm 0cm 1cm}, clip]{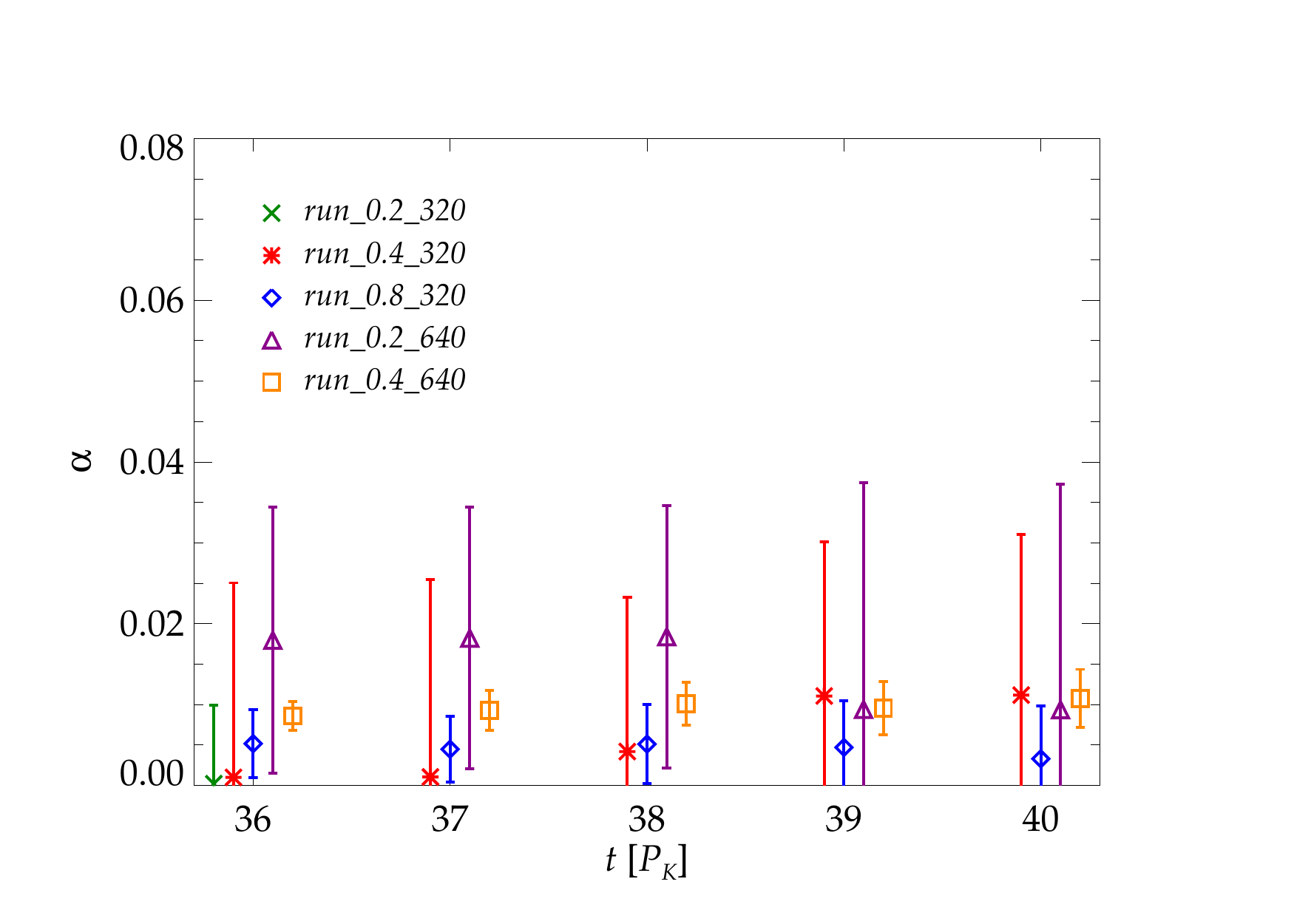}
\includegraphics[width=\textwidth, trim = {0cm 0cm 0cm 1cm}, clip]{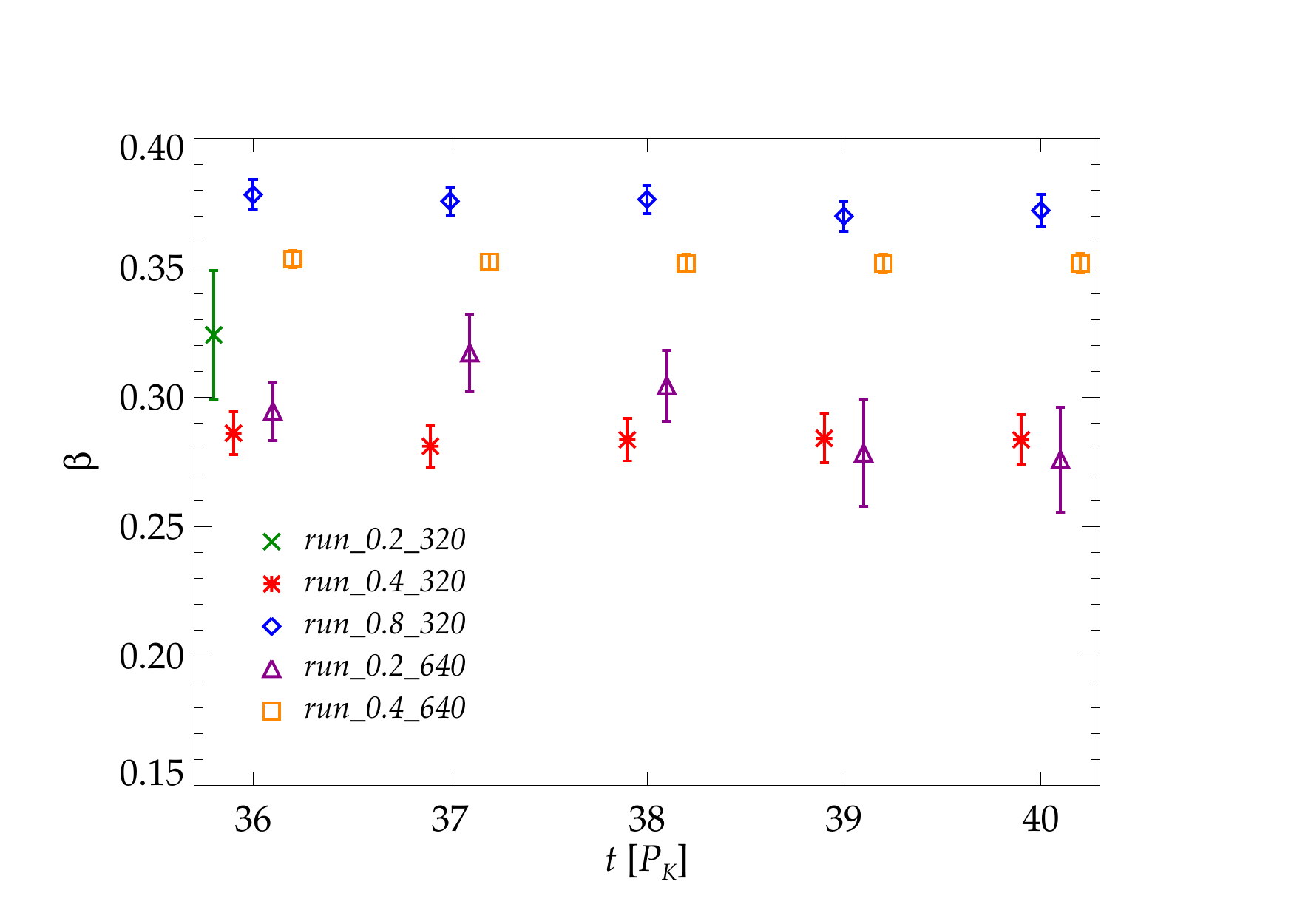}
\end{minipage}
\caption{Best-fitting parameters $M_{\rm{min}}$ (upper left panel), $\alpha$ (upper right panel), $M_{\rm{exp}}$ (lower left panel), and $\beta$ (lower right panel) of the exponentially tapered power law (Eq.~(\ref{exponentially tapered power law M_min})) at every Keplerian orbital period between \mbox{$t=36~P_{\rm{K}}$} and \mbox{$t=40~P_{\rm{K}}$} for all five simulations. Standard errors are plotted as error bars. In the case of the simulation with the smallest box size and the lower resolution, \textit{run\_0.2\_320}, only parameter values for \mbox{$t=36~P_{\rm{K}}$} are plotted because afterwards only four sink particles persist, which prevents us from properly fitting the mass distributions with a power law and an exponential tapering.}
\label{IMF parameters}
\end{figure*}

\begin{table}[t]
\caption{Best-fitting parameters}
\label{best-fitting parameters}
\resizebox{\hsize}{!}{
\begin{tabular}{lccccc}
\hline
\hline
Name&$M_{\rm{min}}~[G^{-1}~H_{\rm{g}}^3P_{\rm{K}}^{-2}]$\tablefootmark{a}&$\langle\alpha\rangle$\tablefootmark{b}&$\langle M_{\rm{exp}}\rangle~[G^{-1}~H_{\rm{g}}^3P_{\rm{K}}^{-2}]$\tablefootmark{b}&$\langle\beta\rangle$\tablefootmark{b}&$N_{\rm{f}}$\tablefootmark{c}\\
\hline
\textit{run\_0.2\_320}&$(1.9\pm3.8)\times10^{-7}$&$0.0002\pm0.0097$&$(1.85\pm0.32)\times10^{-4}$&$0.324\pm0.025$&1\\
\hline
\textit{run\_0.4\_320}&$(1.9\pm0.3)\times10^{-5}$&$0.0057\pm0.0094$&$(2.31\pm0.14)\times10^{-5}$&$0.284\pm0.004$&3\\
\hline
\textit{run\_0.8\_320}&$(1.5\pm0.2)\times10^{-5}$&$0.0045\pm0.0022$&$(1.32\pm0.02)\times10^{-4}$&$0.375\pm0.003$&4\\
\hline
\textit{run\_0.2\_640}&$(4.7\pm2.8)\times10^{-6}$&$0.0147\pm0.0085$&$(5.05\pm0.41)\times10^{-5}$&$0.294\pm0.007$&1\\
\hline
\textit{run\_0.4\_640}&$(1.2\pm0.1)\times10^{-6}$&$0.0097\pm0.0011$&$(1.82\pm0.02)\times10^{-5}$&$0.352\pm0.002$&3\\
\end{tabular}
}
\tablefoot{
Listed errors are standard errors.\\
\tablefootmark{a}{Best-fitting values at  \mbox{$t=36~P_{\rm{K}}$} in the case of \textit{run\_0.2\_320} and at \mbox{$t=40~P_{\rm{K}}$} in the case of the other simulations.}
\tablefoottext{b}{Mean values, averaged over \mbox{$36~P_{\rm{K}}\leq t \leq 40~P_{\rm{K}}$}.}
\tablefoottext{c}{Number of filaments.}
}
\end{table}

\begin{table}[t]
\caption{Sink particle statistics}
\label{statistics}
\resizebox{\hsize}{!}{
\begin{tabular}{lcccc}
\hline
\hline
Name&$N_{\rm{tot}}$&$M_{\rm{min}}~[G^{-1}~H_{\rm{g}}^3P_{\rm{K}}^{-2}]$&$M_{\rm{max}}~[G^{-1}~H_{\rm{g}}^3P_{\rm{K}}^{-2}]$&$\langle M\rangle ~[G^{-1}~H_{\rm{g}}^3P_{\rm{K}}^{-2}]$\\
\hline
\textit{run\_0.2\_320}&4&$5.9\times10^{-5}$&$1.6\times10^{-3}$&$4.8\times10^{-4}$\\
\hline
\textit{run\_0.4\_320}&15&$7.8\times10^{-6}$&$4.3\times10^{-3}$&$4.9\times10^{-4}$\\
\hline
\textit{run\_0.8\_320}&42&$8.3\times10^{-6}$&$9.3\times10^{-3}$&$6.9\times10^{-4}$\\
\hline
\textit{run\_0.2\_640}&7&$1.0\times10^{-5}$&$1.5\times10^{-3}$&$2.6\times10^{-4}$\\
\hline
\textit{run\_0.4\_640}&62&$1.9\times10^{-6}$&$2.2\times10^{-3}$&$1.2\times10^{-4}$\\
\end{tabular}
}
\end{table}

In Fig.~\ref{IMF parameters} the best-fitting parameters $M_{\rm{min}}$, $\alpha$, $M_{\rm{exp}}$, and $\beta$ in Eq.~(\ref{exponentially tapered power law M_min}) are shown for \mbox{$t=36~P_{\rm{K}}$} to \mbox{$t=40~P_{\rm{K}}$} for all of our simulations. In the case of the simulation with the smallest box dimensions and the lower resolution, \textit{run\_0.2\_320}, the small number of sink particles persisting after \mbox{$t=36~P_{\rm{K}}$} does not permit to properly fit a power law and an exponential cutoff. The parameters $\alpha$, $M_{\rm{exp}}$, and $\beta$ are approximately constant in time for all simulations apart from $M_{\rm{exp}}$ for the simulation with the middle box dimensions and the lower resolution, \textit{run\_0.4\_320}, which seems to converge to an upper limit. In Table~\ref{best-fitting parameters}, we list mean values of these three parameters, averaged over the simulation time span from \mbox{$t=36~P_{\rm{K}}$} to \mbox{$t=40~P_{\rm{K}}$}.

Similar to $M_{\rm{exp}}$ for \textit{run\_0.4\_320}, the fitted minimum mass $M_{\rm{min}}$ increases with time for the simulation with the largest box size, \textit{run\_0.8\_320}, and the one with the middle box size and the higher resolution, \textit{run\_0.4\_640}, but appears to saturate at an upper limit. The increase of $M_{\rm{min}}$ and $M_{\rm{exp}}$ is likely due to both pebble accretion by the sink particles -- as less and less pebbles remain towards the end of the simulations (see Fig.~\ref{rhopmx}), both masses converge to an upper limit -- and possibly artificial merging of sink particles. In contrast to this, for \textit{run\_0.4\_320} and the simulation with the smallest box dimensions and the higher resolution, \textit{run\_0.2\_640}, $M_{\rm{min}}$ is roughly constant. To provide a comparable value for each simulation despite these differences in the time dependence, in Table~\ref{best-fitting parameters} we list the best-fitting values at \mbox{$t=40~P_{\rm{K}}$} for all simulations (except for \textit{run\_0.2\_320}, for which only the value at \mbox{$t=36~P_{\rm{K}}$} is available). A comparison of these values shows that $M_{\rm{min}}$ declines with both increasing box dimensions and increasing resolution if the value for \textit{run\_0.2\_320} is disregarded.

The best-fitting values of the exponent of the power-law component $\alpha$ for all simulations vanish. This is because a lower limit for the sizes of the pebble clusters, and thus the sink particle masses, is set by the resolution, which even in the case of the higher resolution is too large for low-mass sink particles that would constitute the power-law part to emerge. Both \citet{Johansen2015} and \citet{Simon2016} fit the differential mass distribution with a power-law exponent of about~$-1.6$, corresponding to \mbox{$\alpha=0.6$}. Since the higher resolution we employ, $640~H_{\rm{g}}^{-1}$, is the lowest one that is considered in these papers, to properly study the power-law distribution higher resolutions than $640~H_{\rm{g}}^{-1}$ seem to be required.

We find the mass scale of the exponential cutoff $M_{\rm{exp}}$ to correlate with the mass budget in every filament. The parameter $M_{\rm{exp}}$ should increase with the distance between the filaments because, if the distance is larger, more pebbles can be accreted by the planetesimals forming in each filament. In Col.~6 of Table~\ref{best-fitting parameters}, the numbers of filaments $N_{\rm{f}}$ we observe in our simulations are listed. One, three, and four filaments form in the simulation boxes with radial and azimuthal dimensions of $0.2~H_{\rm{g}}$, $0.4~H_{\rm{g}}$, and $0.8~H_{\rm{g}}$, respectively.  Therefore, the mass reservoir of pebbles in each filament is similar for the smallest and the largest box sizes, but smaller for the middle box sizes.  We indeed see that the mean values of $M_{\rm{exp}}$ for \textit{run\_0.2\_320} and \textit{run\_0.8\_320} are similar, but larger than the one for \textit{run\_0.4\_320}.  Likewise, the mean value for \textit{run\_0.2\_640} is greater than that for \textit{run\_0.4\_640}. Even though the mean values for the two simulations with the middle box size differ by more than one standard deviation, we note that the best-fitting values for \textit{run\_0.4\_320} increase with time with a range enclosing the mean value for \textit{run\_0.4\_640}. Hence, we find $M_{\rm{exp}}$ to be largely independent of the resolution.

With the mean values of the exponent of the exponential cutoff $\beta$ ranging from 0.28 to 0.38, the exponential cutoff is rather smooth. \citet{Johansen2015} fit their data by eye using \mbox{$\beta=4/3$}, which is a significantly steeper cutoff, but this steep cutoff might be an artefact of the small box size they employed, in which only a few massive planetesimals formed. The best-fitting values for the two simulations with the largest number of sink particles, \textit{run\_0.8\_320} and \textit{run\_0.4\_640}, are nearly equal, but somewhat greater than the values for \textit{run\_0.4\_320} and \textit{run\_0.2\_640}, which are also roughly equal. This indicates that only in the former two simulations enough sink particles emerge to completely capture the high-mass end of the mass distribution. However, the mean values for all simulations lie in a rather small range of 0.1, hence we find $\beta$ to be relatively independent of the box size and the resolution.

Substituting the best-fitting parameters listed in Table~\ref{best-fitting parameters} into the cumulative (Eq.~(\ref{exponentially tapered power law M_min})) or the differential mass distribution,
\begin{equation}
\begin{split}
\frac{dN}{dM}=\frac{1}{M}~\left[\alpha+\beta\left(\frac{M}{M_{\rm{exp}}}\right)^{\beta}\right]~\left(\frac{M}{M_{\rm{pow}}}\right)^{-\alpha}\\
\times\exp\left[\left(\frac{M_{\rm{min}}}{M_{\rm{exp}}}\right)^{\beta}-\left(\frac{M}{M_{\rm{exp}}}\right)^{\beta}\right],
\end{split}
\label{differential mass distribution}
\end{equation}
yields an initial mass function for each simulation. The cumulative mass distribution can be converted to a cumulative size distribution,
\begin{equation}
\frac{N_>(R)}{N_{\rm{tot}}}=\left(\frac{R}{R_{\rm{min}}}\right)^{-3\alpha}~\exp\left[\left(\frac{R_{\rm{min}}}{R_{\rm{exp}}}\right)^{3\beta}-\left(\frac{R}{R_{\rm{exp}}}\right)^{3\beta}\right],
\end{equation}
where $R$ is the radius of every sink particle and the minimum radius $R_{\rm{min}}$ and the radius scale of the exponential tapering $R_{\rm{exp}}$ can be calculated from $M_{\rm{min}}$ and $M_{\rm{exp}}$, respectively, using a solid density of~$3~\rm{g}~\rm{cm}^{-3}$.

In Table~\ref{statistics}, we list the number $N_{\rm{tot}}$, minimum mass $M_{\rm{min}}$, maximum mass $M_{\rm{max}}$, and mean mass $\langle M\rangle $ of the sink particles at the end of our five simulations, \mbox{$t=40~P_{\rm{K}}$}. In the case of the two simulations with the smallest box dimensions, we observe less than ten sink particles, and the best-fitting parameters for these simulations are therefore afflicted with comparably large errors (see Fig.~\ref{IMF parameters}). We find the actual $M_{\rm{min}}$ and $\langle M\rangle $ to be of the order of the fitted $M_{\rm{min}}$ and $M_{\rm{exp}}$, respectively (compare with Cols.~2 and 4 of Table~\ref{best-fitting parameters}). For \textit{run\_0.8\_320} and \textit{run\_0.4\_640}, the 10\% of the sink particles which are most massive contain 66\% and 70\%, respectively, of the total sink particle mass. That is, in our simulations the most massive sink particles dominate the total mass. 

As stated above, a higher resolution enables us to observe the formation of smaller pebble clusters, and thus less massive sink particles. Hence, $N_{\rm{tot}}$ increases and both $M_{\rm{min}}$ and $\langle M\rangle$ decline with increasing resolution. Like \citet{Johansen2015} and \citet{Simon2016}, we find the maximum mass $M_{\rm{max}}$ to be relatively independent of the resolution. 

We expect the number of planetesimals to increase with the number of filaments, and thus with the radial box dimension, and with the length of the filaments, i.e.\ the azimuthal box dimension. Analogously to the mass scale of the exponential cutoff $M_{\rm{exp}}$, as discussed above, we further expect $\langle M\rangle$ to increase with the distance between the filaments. Our findings are consistent with these expectations, with the exception of $\langle M\rangle$ for \textit{run\_0.2\_320}. One and three filaments form in \textit{run\_0.2\_320} and \textit{run\_0.4\_320}, respectively (see Col.~6 of Table~\ref{best-fitting parameters}), therefore the value for the former simulation should be greater than the one for the latter simulation, yet it is slightly smaller. This shows that, at least at the lower resolution, the smallest boxes, in which only one filament forms, might be too small to accurately capture the mass budget of each filament. 

Furthermore, $M_{\rm{min}}$ and $M_{\rm{max}}$ in general increase with the box dimensions, but we find $M_{\rm{min}}$ to considerably decrease if the box size is increased from $0.2~H_{\rm{g}}$ to $0.4~H_{\rm{g}}$ in the radial and azimuthal directions. This also indicates that the smallest box dimensions might not capture the sink particle mass distribution as well as the larger boxes. However, it may also be a stochastic effect because, especially for a small number of sink particles, the ensemble-averaged values of $M_{\rm{min}}$ and $M_{\rm{max}}$ might differ significantly from the actual values.

\section{Summary and discussion}
We have investigated the formation of planetesimals by the streaming instability in numerical simulations with three different box sizes and two different resolutions. In particular, we have studied the initial mass function of these planetesimals, employing the largest box dimensions to date with radial and azimuthal sizes of up to 0.8~gas scale heights. These large box sizes have enabled us to study planetesimal formation in multiple axisymmetric filaments formed by the streaming instability and to yield better statistics because more planetesimals emerge in simulations with larger box sizes. 

In the absence of self-gravity, the streaming instability concentrates pebbles into axisymmetric filaments. After self-gravity has been introduced, these filaments disperse within about ten Keplerian orbital periods because the pebbles accumulate into clusters that undergo gravitational collapse and form planetesimals. We have observed that, after their formation, the planetesimals on average migrate through more than half of the radial dimension of the simulation box owing to mutual gravitational scattering. The extent of the radial migration does not converge for the box sizes we have taken into consideration. Further studies could provide insights regarding the implications of the migration through multiple filaments for the dependence of the composition of planetesimals on the orbital distance.

The radii of the planetesimals formed in our simulations, which depend on the strength of the self-gravity and thus on the solid particle column density, range from 80~km to 620~km. We have compared power-law fits to their cumulative mass distribution with and without exponential tapering and have found that a rather shallow exponential cutoff fits the distribution better than the steeper cutoff of an integrated power law. \citet{Johansen2015} also find the initial mass function to be represented best by an exponentially tapered power law, although they studied planetesimals that are smaller than the ones formed in our simulations. In their simulation with the highest resolution, the planetesimal radii amount to between 30~km and 120~km. In contrast to this, \citet{Simon2016} find that a power law without exponential tapering is suitable to fit the birth mass distribution of planetesimals with radii between 50~km and a few hundred kilometers.

We have found a value of the exponent of the exponential cutoff of about 0.3 to 0.4, which is largely invariant under changes in the box size and the resolution, but considerably smaller than the value of~4/3 that \citet{Johansen2015} determine, which is based on a much smaller number of massive planetesimals that formed in a small simulation domain. However, the resolutions we have considered are insufficient to constrain the shape of the power-law part and to investigate its dependence on the box dimensions because the planetesimals that formed in our simulations are too large to constitute a power-law distribution at the low-mass end of the initial mass function.

Both the characteristic mass of the exponential cutoff and the mean mass of the planetesimals correlate with the pebble mass budget in every filament. In this regard, we have found indications that a simulation box with a size of 0.2~gas scale heights in the radial, azimuthal, and vertical directions, in which only one filament emerges, may be too small to properly capture the mass reservoir. This is consistent with the observation by \citet{Yang2014} that box dimensions of 0.2~scale heights in the radial and azimuthal directions are too small to capture all scales relevant for the streaming instability, and is further supported by Li et al.'s (in prep.) finding that the density distribution function of solid particles is consistent only for boxes with radial and azimuthal sizes of at least 0.4~scale heights.

The current size distribution of the asteroids in the asteroid belt with diameters between 120~km and several hundred kilometers, corresponding to the sizes of the planetesimals that emerge in our simulations, is well-fitted with a power law. This power law, which \citet{Bottke2005} argue represents the primordial asteroid size distribution, is in contrast to the exponential cutoff we (and \citealt{Johansen2015}) find. Subsequent pebble accretion therefore appears to be necessary to convert the exponential tapering of the birth size distribution into the power law observed in the asteroid belt \citep{Johansen2015}.

It is interesting to compare the initial mass function of planetesimals to the classical concept of an initial mass function of stars. The formation of stars is comparable to the formation of planetesimals by the streaming instability insofar as both stars and planetesimals form by gravitational collapse, the former from molecular cloud cores, and the latter from pebble clusters. The differential mass distribution of stars with masses greater than about $1~M_{\sun}$ is given by a power law with an exponent of approximately $-2.3$ \citep{Salpeter1955, Massey1998, Chabrier2003}. That is, the total stellar mass is dominated by the least massive stars, in contrast to the total mass of the planetesimals we have observed in our simulations, which is dominated by the most massive ones. It has been argued that there is a physical upper mass cutoff of the stellar initial mass function, but massive stars are rare and short-lived, and their mass distribution is therefore difficult to observe \citep{Zinnecker2007}.

\citet{Johansen2015} and \citet{Simon2016} investigated the dependence of the shape of the planetesimal initial mass function on the resolution, the pebble column density, the strength of the self-gravity, and the simulation at which self-gravity is initiated. We have complemented these parameter studies with an analysis of the box-size dependence, but how, for instance, the solid-to-gas ratio, the friction time of the pebbles, the radial gas pressure gradient, and the vertical box size affect the shape of the birth mass distribution remains to be investigated. Finally, the streaming instability has been shown to operate in protoplanetary disks with turbulence driven by the magnetorotational instability \citep{Johansen2007b}. It remains to be seen how turbulence and magnetic fields influence the shape of the initial mass function.

\begin{acknowledgements}
We thank Piero Ranalli for his advice on how to calculate the standard errors of the fitting parameters. We further thank the anonymous referee for providing comments and questions that helped to improve the paper. The simulations presented in this paper were performed on resources provided by the Swedish National Infrastructure for Computing (SNIC), the Alarik system at Lunarc at Lund University and the Beskow system at the PDC Center for High Performance Computing at the KTH Royal Institute of Technology. This research was supported by the European Research Council under ERC Starting Grant agreement 278675-PEBBLE2PLANET. A.\ J.\ is thankful for financial support from the Knut and Alice Wallenberg Foundation and from the Swedish Research Council (grant 2014-5775).
\end{acknowledgements}

\bibliography{Initial_mass_function_of_planetesimals_formed_by_the_streaming_instability.bib} 

\end{document}